\documentclass[10pt,twocolumn,preprintnumbers,amsmath,amssymb,nofootinbib
,superscriptaddress]{revtex4-1}
\usepackage{graphicx,longtable,mathrsfs,color,array}
\usepackage[hidelinks]{hyperref}
\usepackage[usenames,dvipsnames]{xcolor} % colour
\usepackage{amssymb,amsmath,mathtools,mathrsfs,slashed} % maths
\usepackage{epsfig,subfigure,placeins,float} % plots
\usepackage{booktabs,longtable,ctable,multirow} % tables
\usepackage{exscale,relsize} % scale 
\usepackage[normalem]{ulem} % editing
\usepackage{enumerate}
\usepackage{times, mathptmx} % Times font
\usepackage{comment}
\usepackage{color}
\allowdisplaybreaks[1]

 \definecolor{DarkBlue}{cmyk}{1,1,0,0.1}

\begin{document}
\title{Curvature Perturbations in the Effective Field Theory of Inflation}
\author{Macarena Lagos}
\email{mlagos@kicp.uchicago.edu}
\affiliation{Kavli Institute for Cosmological Physics, The University of Chicago, Chicago, IL 60637, USA}
\author{Meng-Xiang Lin}
\email{mxlin@uchicago.edu}
\affiliation{Kavli Institute for Cosmological Physics, The University of Chicago, Chicago, IL 60637, USA}
\affiliation{Department  of  Astronomy \& Astrophysics, Enrico Fermi Institute, The University of Chicago, Chicago, IL 60637, USA}
\author{Wayne Hu}
\email{whu@background.uchicago.edu}
\affiliation{Kavli Institute for Cosmological Physics, The University of Chicago, Chicago, IL 60637, USA}
\affiliation{Department  of  Astronomy \& Astrophysics, Enrico Fermi Institute, The University of Chicago, Chicago, IL 60637, USA}
\date{Received \today; published -- 00, 0000}

\begin{abstract}
We discuss the difference between various gauge-invariant quantities typically used in single-field inflation, namely synchronous $\zeta_s$, comoving $\zeta_c$, and unitary $\zeta_u$ curvatures. 
We show that conservation of $\zeta_c$ outside the horizon is quite restrictive on models as it leads to conservation of $\zeta_s$ and $\zeta_u$, whereas the reverse does not hold.
We illustrate the consequence of these differences with two inflationary models: ultra-slow-roll (USR) and braiding-ultra-slow-roll (BUSR). In USR, we show that out of the three curvatures, only $\zeta_s$ is conserved outside the horizon, and we connect this result to the concepts of separate universe and the usage of the $\delta N$ formalism. We find that even though $\zeta_s$ is conserved, there is still a mild violation of the separate universe approximation in the continuity equation.
Nevertheless, the $\delta N$ formalism can still be applied to calculate the primordial power spectrum of some gauge-invariant quantities such as $\zeta_u$, although it breaks down for others such as the uniform-density curvature.
In BUSR, we show that both $\zeta_u$ and $\zeta_s$ are conserved outside the horizon, but take different values. Additionally, since $\zeta_u\not=\zeta_c$ we find that the prediction for observable curvature fluctuations after inflation does not reflect $\zeta_c$ at horizon crossing during inflation and moreover involves not just $\zeta_u$ at that epoch but  also the manner in which the braiding phase ends.  
\end{abstract}

\date{\today}
\maketitle

%%%%%%%%%%%%%%%%%%%%%%%%%%%%%%%%%%%%%%%%%%%%%%%%%%%%%%%%%%%%%%%%%%%%%%%%%%%%%%%%%%%%%%%%%%%%%%%%%%%%%%%%%%%%%%%%%%%%%%%%%%%%%%%%%%%%%%%%%%%%%%%%%%%%%%

\section{Introduction}\label{sec:introduction}
Observations of the Cosmic Microwave Background temperature anisotropies \cite{Aghanim:2018eyx,Akrami:2018odb} are in excellent agreement with an early universe with primordial perturbations that are adiabatic and have a nearly scale-invariant power spectrum. Currently, the most compelling explanation for these initial perturbations is the inflationary paradigm, where the universe expanded nearly exponentially fast and matter perturbations were seeded by the vacuum fluctuations of one or more fields. The simplest scenario is described by a single scalar field that slowly rolls down a potential well towards the end of inflation, leading afterwards to reheating and the production of the Standard Model particles.

Predictions on the power spectra and bispectra of inflationary models can be obtained using different gauge-invariant variables, such as the the unitary curvature $\zeta_u$, comoving curvature $\zeta_c$,  synchronous curvature $\zeta_s$, or uniform-density curvature $\zeta_\rho$, among others. The usage of each one of these variables has its own advantages. For instance, unitary curvature describes the spatial curvature of the spacetime in a frame where the scalar field evolution provides a clock that breaks temporal but preserves spatial diffeomorphism invariance. In this frame, it is easy to generically describe single-field inflationary models with the Effective Field Theory of inflation \cite{Cheung:2007st}, where the only explicit perturbation fields come from the spacetime metric. In addition, synchronous curvature is useful for the concept of a local background or separate universe \cite{Hu:2016ssz}, which can be used to straightforwardly estimate the effects of long-wavelength perturbations on the local universe.  The separate universe concept itself is closely related to the $\delta N$ technique which is often used to calculate non-Gaussianity from inflation using the $e$-folding of a local background  \cite{Sasaki:1995aw, Wands:2000dp, Sugiyama:2012tj,Domenech:2016zxn,Abolhasani:2018gyz,doi:10.1142/10953}.
Finally, comoving curvature and uniform-density curvature are common variables used to connect inflationary fluctuations to observables.

In typical slow-roll inflationary models found in the literature, all the previous variables mentioned above either coincide or have the same qualitative behaviour, and therefore they all provide the same information and we can simply choose the most convenient one. However, in general single-field models this is not the case. In this paper, we exploit this difference and analyze one of the main features of typical inflationary models---conservation of curvature outside the sound horizon---for $\zeta_u$, $\zeta_c$ and $\zeta_s$. We find that, in general, conservation of $\zeta_c$ is the most restrictive condition as it will imply conservation of both $\zeta_s$ and $\zeta_u$, but the reverse will not hold. In addition, we identify sufficient conditions on inflationary models in order to have conserved curvatures given that either $\zeta_u$ or $\zeta_s$ is conserved.

In order to illustrate the difference between these three curvatures, we explicitly discuss two inflationary models. First, we consider ultra-slow-roll (USR) inflation \cite{Kinney:2005vj}, in which $\zeta_c=\zeta_u$ grows in time outside the horizon, but $\zeta_s$ is conserved. Second, we build a new model dubbed braiding ultra-slow-roll (BUSR) inflation, in which $\zeta_u$ and $\zeta_s$ are conserved (but different) whereas $\zeta_c$ grows outside the horizon. 

Furthermore, we discuss some conceptual and observational consequences of having different curvatures for USR and BUSR. First, we discuss the concept of separate universe, where super-horizon
perturbations can be reabsorbed into the background equations such that the total perturbed universe still looks homogeneous and isotropic in a local Hubble-sized patch \cite{Wands:2000dp}. As shown in \cite{Hu:2016ssz, Hu:2016wfa}, separate universe is valid when synchronous observers see a local approximate FRW universe, which requires conservation of $\zeta_s$ outside the horizon. Here we show that USR does have $\zeta_s$ conserved but still violates separate universe via the continuity equation. In addition, we discuss the $\delta N$ formalism. 
This formalism is typically assumed to require the validity of separate universe, although here we show that it can still be used in USR for obtaining observables in terms of appropriate variables, but subtleties can arise when using the formalism for other variables. 
In particular, we find that the standard $\delta N$ prescription (see \cite{Lyth:2004gb}) yields the correct value for $\zeta_u$ but the incorrect value for $\zeta_\rho$ in USR.

Finally, we explore the observational consequence of BUSR in detail. In  general, inflationary models provide initial conditions for the matter distribution of the universe during early times, which are typically in turn provided by the value of $\zeta_u$ at horizon crossing. This value then determines the value of $\zeta_c(t_\text{end})$ at the end of inflation, which is then propagated forward to radiation and matter domination outside the horizon, ultimately becoming the initial
condition for structure formation. This translation  between unitary and comoving curvature is straightforwardly done when both curvatures are conserved and take the same value 
outside the horizon. However, in BUSR we have that $\zeta_c\not=\zeta_u$ due to a non-trivial coupling called braiding, which corresponds to derivative interactions between the metric and the inflationary field. In this paper, we analyze the difference between $\zeta_u$ at horizon crossing and $\zeta_c(t_\text{end})$ in BUSR. We consider a realistic scenario where braiding vanishes before the end of inflation in order to avoid spoiling the subsequent reheating process \cite{Ohashi:2012wf,Lopez:2019wmr}. In this case, we will have that $\zeta_u (t_\text{end})=\zeta_c(t_\text{end})$, however for some wavelengths $\zeta_u$ evolves outside the horizon and hence $\zeta_c(t_\text{end})$ will not be given by $\zeta_u$ at horizon crossing. In particular, we find that for sufficiently superhorizon perturbations at the time the braiding vanishes, $\zeta_u$ remains frozen and its value at horizon crossing becomes $\zeta_c(t_\text{end})$. Meanwhile, for wavelengths that have only been outside the horizon for a few $e$-folds before this epoch, $\zeta_u$ evolves, which ultimately leads to a suppression of the power spectrum at the end of inflation for these and smaller scales.

This paper is structured as follows. In Section \ref{sec:setup} we give a general fluid description for generic inflationary models, then we introduce some relevant gauge-invariant fields that are typically used for inflation, and finally we review the concept of separate universe. 
In Section \ref{sec:unitary} we discuss the general relationship between the conservation of the curvature in unitary, comoving, and synchronous gauge in single-field inflationary models. 
In Section \ref{sec:examples} we give examples of inflationary models that illustrate the differences in curvatures discussed in the previous section.
In Section \ref{sec:observables} we discuss the conceptual and observational consequences of having curvatures evolving differently for the models presented in \ref{sec:examples}, including the consequence for separate universe, the $\delta N$ formalism, and the evolution of curvatures outside the horizon.
Finally, in Section \ref{sec:discussion} we summarize our results and discuss their relevance. Throughout this paper we will be using Planck units, with $c=1$ and $8\pi G=1$.

\section{General Description}\label{sec:setup}

\subsection{Effective Fluid Decomposition}
Let us start by considering an inflationary model in a spatially-flat cosmological FRW background, with small scalar perturbations. In this case we can write the metric as:
\begin{align}\label{MetricPerts}
&ds^2=-(1+2\Phi)dt^2+2B_{,i} dx^i dt + a(t)^2\left [(1-2\Psi)\delta_{ij}\right. \nonumber\\
& \left. +2E_{,ij}\right]dx^idx^j,
\end{align}
where subscript commas denote derivatives throughout, $a(t)$ is the background scale factor whereas $\Phi$, $\Psi$, $B$ and $E$ are the four metric perturbations. We shall often refer to $\Psi$ as the curvature perturbation in the $3+1$ slicing defined by the lapse perturbation $\Phi$ and the shift $B_{,i}$. All these perturbation fields depend on space and time. Even though there may be non-trivial interactions between the fields driving inflation and the metric, we can always write the equations of motion in an Einstein-like form:
\begin{equation}\label{EinsEqns}
G^{\mu}{}_\nu=T^{\mu}{}_\nu,
\end{equation} 
where $G^{\mu}{}_\nu$ is the Einstein tensor, and $T^{\mu}{}_\nu$ is an effective stress tensor from all the possible inflationary fields. Similarly to eq.~(\ref{MetricPerts}) we can decompose $T^{\mu}{}_\nu$ into a background and perturbative part in the following fluid-like way:
\begin{align}\label{StressPerts}
& T^0{}_{0}= \rho+\delta \rho, \nonumber\\
& T^i{}_{0}=-(\rho+p)\bar{g}^{ij}\partial_j (v-B),\nonumber\\
&T^0{}_{i}=(\rho+p)\partial_i v, \nonumber\\
& T^i{}_{j}=\left(p+\delta p\right)\delta^i{}_j -p\left( \bar{g}^{il}\partial_l\partial_j -\frac{1}{3}\delta^i{}_j \bar{g}^{lk}\partial_k\partial_l\right)\pi,
\end{align}
where $\bar{g}_{ij}=a^2\delta_{ij}$ is the spatial background metric. Here we have retained linear terms in the perturbations when raising and lowering indices.
Here, $\rho$ and $p$ are some effective background energy density and pressure. In addition, $\delta \rho$, $v$, $\delta p$ and $\pi$ are the four effective fluid perturbations describing the energy-density, velocity potential, pressure, and anisotropic stress, respectively. 

According to the decomposition given in eq.~(\ref{MetricPerts}) and (\ref{StressPerts}), the background equations will be given by the $(00)$ and trace $(ij)$ components of eq.~(\ref{EinsEqns}):
\begin{align}
&	3H^2=\rho, \label{eqbkg1}\\
& 2\dot{H}=-(\rho+p),\label{eqbkg2}
\end{align}
where $H=\dot{a}/a$ is the Hubble rate, where dots denote derivatives with respect to $t$. Similarly, the perturbed equations of motion in Fourier space will be given by the $(00)$, $(0i)$, and combinations of the trace-free and trace $(ij)$ components:
\begin{align}
&2\frac{k^2}{a^2}\left[\Psi+H(a^2\dot{E}-B)\right]+6H(\dot{\Psi}+H\Phi)=-\delta \rho,\label{eq00}\\
&2\dot{\Psi}+2H\Phi=-(\rho+p)v,\label{eq0i}\\
& \Psi-\Phi+ (a^2\dot{E}-B)\dot{\vphantom{)}}+H(a^2\dot{E}-B)= -p \pi ,\label{EqAnisotropic}\\
& 2\ddot{\Psi}+2H\dot{\Phi}+4\dot{H}\Phi+6H\left(\dot{\Psi}+H\Phi\right)=\delta p +\frac{2}{3}\frac{k^2}{a^2}p\pi.
\end{align}
Here, all the perturbation fields depend on the wavenumber $k$ and time $t$, although this explicit dependence has been omitted.

\subsection{Gauge}

In single-field inflation, temporal diffeomorphism invariance is broken by the time evolution of the
scalar field in the background, leaving a preferred temporal foliation called unitary slicing where the field is 
spatially unperturbed.  The curvature perturbation generated during inflation in this slicing
is the dynamical quantity that controls adiabatic fluctuations after but does not generally correspond to the curvature fluctuations seen by
specific observers whose clocks are synchronized differently.
 To relate observables in other frames, 
 let us consider the following change of coordinates:
\begin{equation}
x^\mu \rightarrow x^\mu+\xi^\mu(x),
\label{gaugetransform}
\end{equation}
where $\xi^\mu$ is an arbitrary infinitesimal function of $x^\mu$. Under eq.~(\ref{gaugetransform}), the linear metric perturbations transform as:
\begin{align}
& \Phi \rightarrow \Phi - \dot{\xi}^0,  \quad \Psi \rightarrow \Psi +  H\xi^0, \nonumber\\
&  E\rightarrow E-\xi, \quad B\rightarrow B+\xi^0-a^2\dot{\xi},
\end{align} 
where we have defined $\xi^i=\partial^i \xi$. Fluid perturbations, including the effective fluid defined in the previous section, transform as:
\begin{alignat}{2}
& \delta \rho \rightarrow \delta\rho -\xi^0\dot{\vphantom{A}\rho} , \quad  && \delta p \rightarrow \delta p -\xi^0 \dot{p} , \nonumber\\
& v \rightarrow v  +\xi^0, \quad && \pi \rightarrow \pi .
\end{alignat}
Similarly, a scalar field like the inflaton $\varphi$ transforms as:
\begin{equation}
 \delta \varphi\rightarrow   \delta \varphi - \dot{\bar{\varphi}}\xi^0,
\end{equation} 
where $\bar{\varphi}(t)$ describes the background value of the scalar-field, and $\delta\varphi$ its linear perturbation.

For single-field inflationary models coupled to a massless spin-2 field metric (as in GR), there will be only one physical scalar degree of freedom propagating which can be taken to be the unitary curvature.
It is convenient to define observables like unitary curvature in terms of gauge-invariant combinations of the variables which are then valid in any gauge.
The gauge-invariant form of the various curvature observables which we shall use below are given by:
\begin{itemize}
\item Unitary Curvature: $\zeta_u=\Psi-Hv_\varphi$,
\item Comoving Curvature: $\zeta_c= \Psi-Hv$,
\item Synchronous Curvature: $\zeta_s= \Psi-Hv_m$,
\item Uniform Density Curvature: $\zeta_{\rho_i} = \Psi + {\delta \rho_i}/\rho_i'$,
\end{itemize}
where the velocity potentials are: $v$ for the total effective fluid, $v_\varphi\equiv -\delta\varphi/\dot{\bar{\varphi}}$  for
the scalar field, and $v_m$ for  non-relativistic test particles that are initially at rest with respect to the background expansion, and $\rho_i$ is the true energy density of some matter species $i$.  
Here and throughout primes denote derivatives with respect to the background $e$-folds, that is, ${}'= d/d\ln a= H^{-1}d/dt$. 
Note that the sign of the curvature fluctuation is opposite to \cite{Bardeen:1980kt} and much of the literature.

All these definitions are constructed in such a way that they describe the spatial curvature perturbation $\Psi$ in a given coordinate system. The unitary curvature $\zeta_u$ describes the spatial curvature as seen by observers that follow the time slicings determined by the perturbations of the scalar field, and thus see $\delta\varphi=0$. The comoving curvature $\zeta_c$ describes the spatial curvature as seen by observers that comove with the total effective fluid velocity and hence see $v=0$. Analogously, $\zeta_s$ describes the curvature seen by observers that trace non-relativistic matter and see $v_m=0$. Finally $\zeta_{\rho_i}$ is the curvature on surfaces of spatially uniform density in some matter species $i$.
Since $\zeta_{\rho_i}$ depends on the matter species in question, and is mainly
used in cases where there are multiple fields, we do not consider it further in this section. Its conservation requires a small non-adiabatic stress
and velocity divergence in the component (see e.g.~\cite{Wands:2000dp}).

The relationships among the remaining three curvatures, for any single-field inflationary model, are given by:
\begin{align}
\zeta_c&=\zeta_u -H(v-v_\varphi), \label{CUcurvature1}\\
\zeta_c&= \zeta_s - H(v-v_m). \label{CScurvature}
\end{align}
In addition, we also use spatially-flat gauge, a time-slicing where the spatial metric fluctuations vanish: $\Psi=E=0$.  The  dynamical field is then given by the scalar field perturbation $\delta \varphi$. This choice is widely used in inflation and, as we will see later, it is particularly useful to calculate the inflationary primordial bispectrum of perturbations using the so-called $\delta N$ formalism. The relationship between the unitary curvature $\zeta_u$ and the scalar field in spatially-flat gauge $\delta\varphi_{f}$ is generically at linear order given by:
\begin{equation}\label{ZetaUSF}
\zeta_u=\frac{\delta \varphi_f}{\bar{\varphi}'}.
\end{equation}

Next, we show that whereas typical calculations for inflation are performed using $\zeta_u$ or $\zeta_c$, the concept of separate universe is defined using $\zeta_s$. 

\subsection{Separate Universe} \label{sec:SeparateUniverse}
In the separate universe approach, each super-Hubble sized region of the universe can be considered as a separate FRW universe, with a different effective matter content but locally homogeneous. In particular, long-wavelength perturbations can be absorbed into the background so that the perturbed equations of motion take the same form as those for an FRW universe, and thus the local effect of long-wavelengths cosmological perturbations reduces to a simple change in the background cosmological parameters. 

We can define a separate universe condition that determines when the local universe looks close enough to an actual FRW universe, so that we can apply the separate universe approach. In order to do that, we define a local scale factor $a_W$, effective density $\rho_W$ and spatial curvature $K_W$ such that the (00) equation in (\ref{EinsEqns}) including the background and linear perturbation contributions takes the form:
\begin{align}
\label{Full00eq}
	G^0{}_0 ={}& \bar{G}^0{}_0+ \delta G^0{}_0\equiv  -3(H_W^2+K_W/a_W^2)   \nonumber\\ 
	={} & -\rho _W=-\left(\rho+\delta\rho\right),
\end{align}
where $W$ denotes an windowed average on scales much smaller than the wavelength of the perturbation. Here, the local Hubble factor is defined as $H_W=d\ln a_W/d\tau$, with $d\tau =(1+\Phi)dt$. Explicitly, $H_W=H+\delta H$, where $(\delta H/H)= -\Psi'-\Phi-\Sigma/3$ is the deviation from the background Hubble factor, with $\Sigma=(k/a)^2(a^2E'-B/H)$ quantifying the effective shear of the perturbed expansion of the universe \cite{Wands:2000dp,Hu:2016wfa}. In addition, the curvature is given by:
\begin{equation}\label{KwDef}
K_W=-\frac{2}{3}k^2\Psi.
\end{equation}
Similarly, we rewrite the trace of the $G^i_j$ equation as:
\begin{align}\label{Fullijeq}
 G^i{}_i -G^0{}_0 &=3\left[-\frac{2}{a_W}\frac{d^2a_W}{d\tau^2} -\frac{2}{3}\left(\frac{k}{a}\right)^2\Phi\right] \nonumber\\
 &= \left(\rho_W+ 3p_W\right),
\end{align}
where we have defined $p_W=p+\delta p$.
Also, we rewrite the matter continuity equation as:
\begin{equation}\label{FullContinuity}
\frac{1}{a_W^3}\frac{d (a_W^3\rho_W)}{d\ln a_W}+3p_W= (\rho+p)x^2Hv,
\end{equation}
where we have defined the horizon-to-wavelength ratio
\begin{equation}
 x \equiv \frac{k}{aH}.
 \label{xDef}
\end{equation}
From eq.~(\ref{KwDef}), (\ref{Fullijeq}) and (\ref{FullContinuity}) we see that for infinitely long wavelengths, i.e.~$k\rightarrow 0$, we expect $K_W\rightarrow 0$, $(k/a)^2\Phi\rightarrow 0$ and $(k/a)^2Hv\rightarrow 0$, and hence the perturbed equations to take the same form as a set of Friedmann equations (\ref{eqbkg1})-(\ref{eqbkg2}). Therefore, the effective universe with an infinitely long-wavelength perturbation would look like a homogeneous and isotropic universe. For finite $k$ though, we can define a precise separate universe condition to be that
freely-falling observers that are initially at rest with respect to the background expansion see an approximate FRW effective universe. This frame definition coincides with our definition of synchronous gauge, where $v_m=0$, which implies that $\Phi_s=0$, since the matter stress-energy tensor conservation equation $\nabla^\mu T_{\mu\nu,m}=0$ for a non-relativistic perfect fluid with $p_m=\delta p_m=0$ gives:
\begin{equation}\label{MomentumConsNonrel}
    Hv'_m=-\Phi
\end{equation}
in any frame. Therefore, for synchronous observers, eq.~(\ref{Fullijeq}) looks exactly like one of the Friedmann equations, whereas eq.~(\ref{Full00eq}) will look like a Friedmann equation if the effective spatial curvature $K_W$ is approximately constant. This places the following condition on $\zeta_s=\Psi_s$:
\begin{equation}\label{SUCurvature}
    \left| (\ln \zeta_s)'\right|\ll 1.
\end{equation}
Analogously, the continuity equation will approximate to one in FRW when the source term in the RHS of eq.~(\ref{FullContinuity}) becomes negligible. A sufficient condition for this would be that:
\begin{equation}
\left| (\rho+p)x^2  Hv_s \right|\ll \left| \delta\rho_s \right|,
\end{equation}
where $v_s$ and $\delta \rho_s$ are the perturbed effective velocity energy density in synchronous coordinates. From the equations of motion (\ref{eq00}) and (\ref{eq0i}), this condition can be rewritten as:
\begin{equation}\label{SUContinuity}
\left|x^2 \zeta_s'  \right|\ll \left|x^2 \zeta_s + 3 \zeta_s' + \Sigma_s \right|.
\end{equation}
As long as there are no cancellations on the RHS of eq.~(\ref{SUContinuity}), this condition will be satisfied if (\ref{SUCurvature}) is satisfied and when $|\Sigma_s|\lesssim |x^2\zeta_s|$. Note that the evolution of $\Sigma_s$ can be determined from eq.~(\ref{EqAnisotropic}), which can be rewritten in synchronous gauge as:
\begin{equation}
     \Sigma_s' + \left(3+ \frac{H'}{H}\right)\Sigma_s =-x^2 \left( \zeta_s+p\pi\right),
\end{equation}
which is sourced by the effective anisotropic stress $\pi$ and $\zeta_s$. For minimally-coupled single-field inflationary models we have that $\pi=0$ and hence $|\Sigma_s|\sim |x^2\zeta_s|$. For standard slow-roll models typically considered in the literature eq.~(\ref{SUContinuity}) will then be automatically satisfied whenever eq.~(\ref{SUCurvature}) holds. However, in Section \ref{sec:usr}, we will mention the Ultra-Slow-Roll model, where even though  $|\Sigma_s|\sim |x^2\zeta_s|$, cancellations occur on the RHS of eq.~(\ref{SUContinuity}). In this case,  eq.~(\ref{SUContinuity}) will not follow from  eq.~(\ref{SUCurvature}), and the source term on the RHS of eq.~(\ref{FullContinuity}) will not become negligible. In general, we will say that the separate universe approach holds when (\ref{SUCurvature}) holds {\it and} the source term on the RHS of eq.~(\ref{FullContinuity}) is negligible. 
 
\section{Curvature Conservation}\label{sec:unitary}
In this section we analyze the difference between the three previously mentioned curvatures for single-field inflation, and we discuss situations where a given curvature observable is conserved whereas others are not.    For concreteness, let us consider the most general 
diffeomorphism-invariant action for a single scalar field coupled to the metric with second-order derivative equations of motion, known as the Horndeski action
 \cite{Horndeski:1974wa,Deffayet:2011gz}:
\begin{equation}
S=\int \!\mathrm{d}^4x \sqrt{-g}\left\{\sum_{i=2}^5{\cal L}_i[\phi,g_{\mu\nu}]\right\},
\end{equation} 
where ${\cal L}_i$ are Lagrangians given by:
\begin{align}
{\cal L}_2={} & G_2,\nonumber\\
{\cal L}_3={} &  -G_3 \Box\varphi, \nonumber \\
{\cal L}_4={} &   G_4 R+G_{4,X}\left[ (\Box \varphi)^2-(\nabla_\mu\nabla_\nu\varphi) (\nabla^\mu\nabla^\nu\varphi)\right]  , \nonumber \\
{\cal L}_5={} & G_5 G_{\mu\nu}\nabla^\mu\nabla^\nu\varphi
-\frac{1}{6}G_{5,X}\big[ (\Box\varphi)^3\nonumber \\ 
&  -3(\nabla^\mu\nabla^\nu\varphi)(\nabla_\mu\nabla_\nu\varphi)\Box\varphi \nonumber\\
& +2(\nabla^\nu\nabla_\mu\varphi) (\nabla^\alpha\nabla_\nu\varphi)(\nabla^\mu\nabla_\alpha \varphi)
\big] ,
\end{align}
with $G_{i}(\varphi,X)$, $i\in (2,5)$ as arbitrary functions of $\varphi$ and the kinetic term $X\equiv-\nabla^\nu\varphi\nabla_\nu\varphi/2$.

During inflation, we can characterize the expansion of the universe through $H(t)$ by defining the slow roll parameters: $\epsilon=-H'/H$, and higher derivatives. The equations determining the evolution of the perturbations will be described by four effective parameters $\{\alpha_M, \alpha_T,\alpha_B,\alpha_K\}$ which depend on time only \cite{Bellini:2014fua}. 
In this case, the effective fluid velocity potential $v$ can differ from $v_\varphi$ when the so-called braiding parameter $\alpha_B$ is
nonzero, according to the following expression:
\begin{equation}\label{HorndeskiVel}
v_\varphi-v=\frac{\alpha_B}{2 H\epsilon}(\Phi+Hv'_\varphi).
\end{equation}
 The braiding parameter is explicitly defined as:
\begin{align}
\alpha_B={} & [\varphi'(XG_{3,X}-G_{4,\varphi}-2XG_{4,\varphi X})\nonumber\\
&+4X(G_{4,X}+2XG_{4,XX}-G_{5,\varphi}-XG_{5,\varphi X})\nonumber\\ &+\varphi'XH^2(3G_{5,X}+2XG_{5,XX})]\nonumber\\
&/[G_4-2XG_{4,X}+XG_{5,\varphi}-\varphi'H^2 X G_{5,X}].\label{Braiding_parameter}
\end{align}
Next, we proceed to rewrite the relationship between the three curvatures for the specific case of Horndeski models.
Combining eq.~(\ref{HorndeskiVel}) (\ref{eq0i}), we rewrite eq.~(\ref{CUcurvature1}) solely in terms of the curvature fields as:
\begin{equation}
\zeta_c=\zeta_u -\frac{\Gamma}{\epsilon}\zeta_u';\quad \Gamma\equiv \frac{\alpha_B}{2-\alpha_B}. \label{CUcurvature}
\end{equation}
Similarly, we also rewrite eq.~(\ref{CScurvature}) solely in terms of curvatures. We use eq.~(\ref{eq0i}) and (\ref{eqbkg2}) to rewrite $v$ in terms of the metric perturbations $\Psi$ and $\Phi$. Combining these equations with (\ref{MomentumConsNonrel}), we find that the difference in velocities $(v-v_m)$ can be rewritten in terms of $\zeta'_s$ and obtain the following expression:
\begin{equation}\label{CScurvature2}
\zeta_c=\zeta_s+\frac{{\zeta}_s'}{\epsilon}.
\end{equation}
We emphasize that whereas the relation given in eq.~(\ref{CScurvature2}) is valid for any single-field model, the one in eq.~(\ref{CUcurvature}) is only valid for Horndeski models, and hence conclusions will be restricted to this class of models.

\subsection{Unitary and Comoving}\label{Sec:UC}
The relation between $\zeta_c$ and $\zeta_u$ is given by eq.~(\ref{CUcurvature}),
where $\alpha_B$ is the braiding parameter that depends on the background, and is non-vanishing when the action has kinetic mixing between the metric and the scalar field (e.g.~when there is $X\Box\varphi$). For minimally coupled scalar fields, there is no braiding and therefore comoving and unitary curvatures coincide. Next, we study the relation between these two curvatures for general inflationary Horndeski models when $\alpha_B\not=0$ in the regime of long-wavelength linear perturbations. 

If $\zeta_u(x)$ is conserved for long-wavelength modes, then we can write \cite{Motohashi:2017gqb}:
\begin{equation}\label{ZetaUConserved}
\frac{{\zeta}_u'}{\zeta_u}=  c_s^2 {\cal O}(x^2),
\end{equation}
where $c_s$ is the sound speed of the scalar field. Note then that for conservation of unitary curvature we require $(c_sx)\ll 1$ instead of just $x\ll 1$.
From eq.~(\ref{CUcurvature}) we obtain that $\zeta_c(x)$ will thus be given by:
\begin{align}
\frac{\zeta_c}{\zeta_u} = 1+ \frac{\Gamma}{\epsilon}c_{s}^2 {\cal O}(x^2) .
\label{ZetaCGamma}
\end{align}
On the one hand, we see that $\zeta_c$ will also be conserved as long as the term $\Gamma=\alpha_B/[(2-\alpha_B)]$ is order one or higher in the slow-roll parameter $\epsilon$. In that case, both curvatures $\zeta_c$ and $\zeta_u$ will freeze to the same value outside the horizon. We note that $|\Gamma/\epsilon |\lesssim 1$ is a sufficient condition to have conservation of $\zeta_c$ but it is not necessary. 

On the other hand, for $\Gamma\sim 1$ or larger this is not the case. For instance, when $|\alpha_B|\gg 1$
eq.~(\ref{ZetaCGamma}) is given by
\begin{equation}\label{CUcurvature2}
\frac{\zeta_c }{\zeta_u} = 1+ \frac{ c_s^2}{\epsilon}{\cal O}(x^2),
\end{equation}
and since $|\epsilon|\ll 1$ during inflation, we expect $\zeta_c$ to freeze out much later than $\zeta_u$. In typical inflationary models, $\epsilon$ grows in time, and then if there is a scale at which $\zeta_u$ freezes, then $\zeta_c$ will also eventually freeze. However, there could be transition regimes in which $\epsilon$ decays in time, in which case the difference between $\zeta_u$ and $\zeta_c$ becomes large, and $\zeta_c$ could even grow in time during this regime. 

Conversely, we can use eq.~(\ref{CUcurvature1}) to study the behaviour of $\zeta_u$ when $\zeta_c$ is conserved. Suppose $\zeta_c(x)$ is conserved for long-wavelength modes, that is:
\begin{equation}\label{ZetacConserved}
\frac{\zeta_c'}{ \zeta_c} = {\cal O}(x^2).
\end{equation}
Note that here and below the order counting in $x$ keeps track of the $k$ dependence, although there is generically a $k$-independent prefactor, typically $c_s$, which we omit for simplicity. In that case, all conclusions will hold generalizing the Hubble horizon to the sound horizon, and we use the terms interchangably where no confusion should arise.
Next, we analyze whether $\zeta_c$ will be conserved as a consequence. In order to do this, we write an explicit expression for $\zeta_u$ as a function of $\zeta_c$ and $\zeta_c'$. 
We start by calculating the difference $(v-v_{\varphi})$ in terms of the comoving curvature $\zeta_c$. This velocity difference is a gauge-invariant quantity but we can make a gauge choice to simplify its calculation. In particular, in comoving gauge $(v- v_{\varphi})=-v_{\varphi,c}$, and $v_{\varphi,c}$ can be obtained from eq.~(\ref{HorndeskiVel}) and (\ref{eq0i}) to be:
\begin{equation}\label{vvarphic}
    v_{\varphi,c}= \frac{1}{u}\int d\ln a \; u \frac{\zeta'_c}{H}, 
\end{equation}
where $u\equiv \exp\left(-2\int d\ln a\,[\epsilon/\alpha_B]\right) $ and therefore, from eq.~(\ref{CUcurvature1}) we find:
\begin{equation}\label{ZetaUCInt}
    \zeta_u= \zeta_c - \frac{H
    }{u}\int d\ln a \; u \frac{\zeta'_c}{H}.
\end{equation}
If comoving curvature is conserved according to eq.~(\ref{ZetacConserved}), then 
\begin{equation}\label{ZetaCZetaUConservation}
\frac{\zeta_u}{\zeta_c} = 1 + {\cal O}(x^2) \times
\begin{cases}
\Gamma/\epsilon, &  |\Gamma/\epsilon| \ll 1 \\
1, & |\Gamma/\epsilon| \gtrsim 1
\end{cases}.
\end{equation} 
Therefore, if $\zeta_c$ is conserved for $x\ll 1$, we conclude that $\zeta_u \approx \zeta_c$, and hence $\zeta_u$ will also be conserved for any $\Gamma$ and $\epsilon$, unlike the converse. 

We note that eq.~(\ref{vvarphic}) is valid up to some integration constant that has been ignored, as it depends on the initial velocity, and we have assumed that all perturbations vanish initially, at least in their Hubble time average.
We also emphasize that the conditions that we have discussed for conservation are sufficient but not necessary.

\subsection{Comoving and Synchronous}\label{Sec:CS}
In this section we study the relationship between comoving and unitary curvatures for long-wavelength perturbations. On the one hand, from eq.~(\ref{CScurvature2}) we have that if $\zeta_s(x)$ is conserved, that is:
\begin{equation}\label{ZetaSConserved}
\frac{\zeta_s'}{\zeta_s} =  {\cal O}(x^2),
\end{equation}
then comoving curvature will be given by:
\begin{equation}\label{ZetaCZetaSConserved}
\frac{\zeta_c}{\zeta_s}=1+\frac{{\cal O}(x^2)}{\epsilon}.
\end{equation}
As a result, as discussed in the previous section, depending on how $\epsilon$ behaves in time, $\zeta_c$ could even grow. Therefore, conservation of $\zeta_s$ does not imply conservation of $\zeta_c$ for $x\ll 1$.

On the other hand, if $\zeta_c$ is conserved, we can use eq.~(\ref{CScurvature}) to calculate the velocity difference $(v_m-v)$ in terms of $\zeta_c$. Similarly to the previous section, even though $(v_m-v)$ is gauge invariant, we can make a gauge choice to simplify calculations. In comoving gauge, we use the momentum conservation equation for non-relativistic matter (\ref{MomentumConsNonrel}) to generically obtain that:
\begin{align}\label{Diffvmv}
      v_m-v &= \int d\ln a \; \frac{\zeta'_c}{H},
\end{align}
so that eq.~(\ref{CScurvature}) implies
\begin{equation}
    \zeta_s = \zeta_c - H \int d\ln a \; \frac{\zeta'_c}{H} .
    \label{ZetasInt}
\end{equation}
From this general expression we see that if comoving curvature is conserved for long-wavelength modes according to eq.~(\ref{ZetacConserved}), then $\zeta_s$ will also be conserved. Therefore, we conclude that conservation of $\zeta_c$ generically implies conservation of $\zeta_s$, unlike the converse. 

Similarly to the previous section, eq.~(\ref{Diffvmv}) and (\ref{ZetasInt}) are valid up to some integration constants that have been ignored as we have assumed that all perturbations vanish initially. 

\subsection{Unitary and Synchronous}
In this section, we study the relationship between unitary and synchronous curvatures for long-wavelength modes. We do this by using the results of the previous two sections. 

On the one hand, if unitary curvature is conserved, then, according to the results of Section \ref{Sec:UC}, comoving curvature will also be conserved as long as $|\Gamma/\epsilon|\lesssim 1$. In this case, according to the results of Section \ref{Sec:CS}, synchronous curvature will also be conserved and its value will coincide with $\zeta_u$. This is a sufficient condition that can be generalized. 
Indeed, combining eq.~(\ref{CUcurvature}) and (\ref{CScurvature2}), we have:
\begin{equation}
    \zeta_s + \frac{\zeta_s'}{\epsilon} =
    \zeta_u - \frac{\Gamma}{\epsilon}\zeta_u'.\label{ZetaUS_relation}
\end{equation}
From here we see that if we assume that unitary curvature is conserved then the leading order time dependence on the right 
hand side is $(\Gamma/\epsilon)c_s^2 {\cal O}(x^2)$.  Notice that unlike for $\zeta_c$, the left hand side 
contains a term ${\zeta_s'}/\epsilon$.   Thus even if $\epsilon$ decays quickly,  $\zeta_s$ will still 
be conserved in the same way as $\zeta_u$ as long as $\Gamma \lesssim 1$. As we will see in the next section, although they are then both constant, the values of $\zeta_s$ and $\zeta_u$ do not necessarily coincide. 

On the other hand, if $\zeta_s$ is conserved then, according to eq.~(\ref{ZetaCZetaSConserved}), $\zeta_c$ will also be conserved if $|1/\epsilon|\ll 1$, and hence $\zeta_u$ will also be conserved as shown in eq.~(\ref{ZetaCZetaUConservation}), regardless of whether the given model has a braiding interaction (large or small) or not. Note that, again, this is a sufficient but not necessary condition to have $\zeta_u$ conserved. 

\begin{figure}[h!]
	\centering
	\includegraphics[scale=1.0]{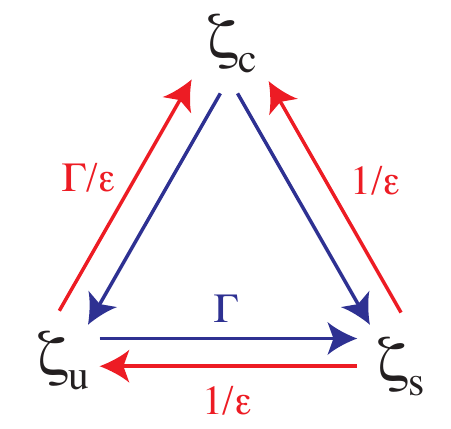}
	\caption{Summary of Section \ref{sec:unitary} on conservation relationships between $\zeta_u$, $\zeta_s$ and $\zeta_c$ for super-horizon linear perturbations in single-field inflation.  Labels on arrows denote additional variables that must be $\lesssim {\cal O}(1)$ for conservation of one curvature to suffice for conservation of the other (see text).}
		\label{Fig:triangle}
\end{figure}

\vspace{1cm}
We summarize the results of this section in Fig.~\ref{Fig:triangle}. Arrows indicate whether conservation of the curvature at the tail suffices to ensure that at the
tip.  
  Labels such as $\Gamma/\epsilon$ indicate any additional
  quantity besides $c_s^2 x^2$ which should be $\lesssim 1$ to establish
  this sufficient condition.

Next, we show examples of inflationary models that illustrate explicitly the inequivalence between the three curvatures studied in this section.

\section{Inflationary models}\label{sec:examples}
In this section we discuss two particular examples of single-field inflationary models that highlight the difference between the three spatial curvatures previously discussed: $\zeta_u$, $\zeta_c$, and $\zeta_s$. In Section \ref{sec:usr} we first describe the ultra-slow-roll model, which exhibits a clear difference in the behaviour of $\zeta_s$ and $\zeta_u=\zeta_c$. In this case, $\zeta_s$ is conserved for super-horizon perturbations, whereas $\zeta_u$ is not conserved and even grows. In Section \ref{sec:busr} we present a new model dubbed braiding-ultra-slow-roll inflation, which has $\zeta_u$ and $\zeta_s$ conserved outside the horizon (although their values differ), whereas $\zeta_c$ grows in time (or even becomes undefined).

In order to find the evolution of linear perturbations in both models, we start by solving the Mukhanov-Sasaki equation for Horndeski models:
\begin{equation}\label{MukhanovSasaki}
    \frac{d^2u}{d\tau^2}+\left(c_s^2k^2-\frac{1}{z}\frac{d^2z}{d\tau^2}\right)u=0,
\end{equation}
where $\tau$ is conformal time related to physical time by $ad\tau=dt$, $u$ describes the physical scalar degree of freedom given by $u\equiv z\zeta_u$, where $z\equiv a\sqrt{2Q_s}$ with $Q_s$ describing an effective Planck mass for scalar perturbations. Also, $c_s$ describes the sound speed of waves for $u$. Explicitly, for Hondeski models with $\alpha_T=\alpha_M=0$ (which will be the case for both USR and BUSR models), the factors $Q_s$ and $c_s^2$ are given by:
\begin{align}
Q_s= &\frac{(2\alpha_K+3\alpha_B^2)}{(2-\alpha_B)^2},\\
c_s^2=& \frac{(2-\alpha_B)(2\epsilon+\alpha_B)+2\alpha_B'}{(2\alpha_K+3\alpha_B^2)},
\end{align}
where $\alpha_B$ is given by eq.~(\ref{Braiding_parameter}) and $\alpha_K$ is defined as:
\begin{align}
\alpha_K=&[(X/H^2)(G_{2,X}+2XG_{2,XX}-2G_{3,\varphi}-2XG_{3,X\varphi})\nonumber\\
&+6\varphi'X(G_{3,X}+XG_{3,XX}-3G_{4,X\varphi}-2XG_{4,XX\varphi})\nonumber\\
&+6X(G_{4,X}+8XG_{4,XX}+4X^2G_{4,XXX})\nonumber\\
& -6X(G_{5,\varphi}+5XG_{5,X\varphi}+2X^2G_{5,XX\varphi})\nonumber\\
&+2\varphi'XH^2(3G_{5,X}+7XG_{5,XX}+2X^2G_{5,XXX})]\nonumber\\
&/[G_4-2XG_{4,X}+XG_{5,\varphi}-\varphi'H^2XG_{5,X}].
\end{align}
The solution to eq.~(\ref{MukhanovSasaki}) for $u$ inside and outside the horizon can be found explicitly for known $z$ and $c_s^2$ functions. Generic solutions have also been discussed in \cite{Motohashi:2017gqb}.

\subsection{Ultra Slow Roll}\label{sec:usr}
Ultra-slow-roll inflation \cite{Kinney:2005vj} is a single-field model, where the scalar field is minimally coupled to the metric, and is one of the few examples that violate the standard consistency relations as the curvature in unitary gauge is not conserved for super-horizon modes \cite{Namjoo:2012aa, Martin:2012pe, Huang:2013lda, Mooij:2015yka, Romano:2016gop, Bravo:2017wyw}. This model has regained attention lately where it has been considered as a transient phase to generate large non-gaussianities and possibly primordial black holes \cite{Germani:2017bcs, Cai:2017bxr, Passaglia:2018ixg}. Explicitly, USR is a canonical scalar field model with:
\begin{equation}
G_{2}(\varphi,X)=X-V(\varphi), \quad G_4(\varphi,X)=\frac{1}{2},   
\end{equation}
and $G_{3}(\varphi,X)=G_5(\varphi,X)=0$, and a potential $V(\varphi)$. The two independent background equations of motion are given by:
\begin{align}
&3H^2=\frac{\dot{\bar{\varphi}}^2}{2}+V(\bar{\varphi}),\label{FriedmanSR} \\
&\ddot{\bar{\varphi}}+3H\dot{\bar{\varphi}}+V_{\bar{\varphi}}=0. \label{ScalarEqSR}
\end{align}
In USR, the potential is driving inflation in eq.~(\ref{FriedmanSR}), but it is extremely flat so that its contribution to eq.~(\ref{ScalarEqSR}) is negligible. The solution to these equations is then such that $\bar{\varphi}'\propto a^{-3}$ and the slow-roll parameter is $\epsilon=\bar{\varphi}'^2/2\propto a^{-6}$. USR then describes an inflationary model that quickly approaches exact de Sitter.

Next, we solve the evolution for perturbations using eq.~(\ref{MukhanovSasaki}). The effective parameters of this model are: $\alpha_K=2\epsilon$, and $\alpha_B=\alpha_T=\alpha_M=0$, and thus $c_s^2=1$ and $Q_s=\epsilon$. The solution for $H$ almost constant and in Bunch-Davies vacuum is given by:
\begin{equation}\label{FullSolu}
    u= \frac{1}{\sqrt{2k}}\left(1+\frac{i}{x}\right)e^{ix}.
\end{equation}
Unitary curvature is then given by 
$\zeta_u=u/(a\bar{\varphi}')$. In the super-horizon limit, where $x \ll 1$, the solution approximates to:
\begin{align}
&\zeta_u \approx  \frac{1}{\bar{\varphi}'}\frac{i H}{\sqrt{2}k^{3/2}} \left[ 1+\frac{1}{2}x^2+ \mathcal{O}(x^3)\right],\nonumber\\
&\zeta_u'\approx \zeta_u\left[3-x^2+ \mathcal{O}(x^3)\right].
\end{align}
We then see that $\zeta_u\propto a^{3}$ and $\zeta_u'\propto a^3$ outside the horizon. Note that since the braiding parameter vanishes in this model, then $\zeta_u=\zeta_c$.

Next, from eq.~(\ref{ZetasInt}) we can analyze the behaviour of synchronous curvature $\zeta_s$. Since both $\zeta_u$ and $\zeta_u'$ grow in the same way in time, both of these terms will contribute equally to $\zeta_s$ and their leading orders in $x$ will actually cancel out. For this reason, it is convenient to rewrite $\zeta_s$ as:
\begin{equation}
\zeta_s= H \int d\ln a \; \epsilon \frac{\zeta_u}{H}.
\end{equation}
This integral can be performed explicitly using the full solution (\ref{FullSolu}), and obtain for super-horizon modes:
\begin{equation}
\zeta_s\approx -8i\epsilon x^{-3}\zeta_u(x=0)\left[ 1 +\mathcal{O}(x^3) \right],
\end{equation}
and therefore its amplitude is approximately constant in time, with $\zeta_s'= \epsilon (\zeta_u-\zeta_s)\approx \epsilon \zeta_u\propto a^{-3}$. We therefore conclude that in USR $\zeta_c=\zeta_u$ grows in time outside the horizon, whereas $\zeta_s$ is conserved.

\subsection{Braiding Ultra Slow Roll}\label{sec:busr}
As previously shown, the difference between $\zeta_u$ and $\zeta_c$ is determined by the quantity $\Gamma/\epsilon$. We will then construct a model that gives a large ratio $\Gamma/\epsilon$, which will hence exemplify a clear case where $\zeta_u\not=\zeta_c$. In order to do this, we use the method described in \cite{Kennedy:2017sof,Kennedy:2018gtx} to construct a single-field model with a de Sitter phase, i.e~with $\epsilon\rightarrow 0$ and a non-vanishing braiding parameter $\alpha_B$. This can be achieved by the following choice of Horndeski functions:
\begin{align}
G_2(\varphi, X)={} &-\left(\Lambda(\varphi)-\frac{15}{8}H_i^2\hat{\alpha}_B\right)\nonumber\\
&+\frac{3}{2}H_i^2\left(-3\hat{\alpha}_{B3}(\varphi)X+\hat{\alpha}_BX^2\right),\nonumber\\
G_3(\varphi,X)={} & H_i\hat{\alpha}_{B3}(\varphi) X,  \quad G_4=\frac{1}{2}, \label{BUSR_G2_Real}
\end{align}
and $G_{5}(\varphi,X)=0$. Here, $\Lambda(\varphi)$ is an arbitrary potential term, $H_i>0$ is an arbitrary parameter which will later determine the Hubble rate of the de Sitter solution, $\hat{\alpha}_B$ is a free constant, and $\hat{\alpha}_{B3}(\varphi)$ is the function that determines the value of the braiding parameter $\alpha_B$. Note that the kinetic term $X$ in $G_2$ is not canonically normalized.

For this model, the background equations of motion are given by:
\begin{align}
%3\left(\frac{H}{H_i}\right)^2 ={} &  \left( \frac{\Lambda(\bar{\varphi})}{H_i^2}-\frac{15}{8}\hat{\alpha}_B\right)  -\frac{9}{4}\hat{\alpha}_{B3}(\bar{\varphi})\dot{\bar{\varphi}}^2 \label{Eq1_bkgd_BUSR}\\
%& +3\frac{H}{H_i} \hat{\alpha}_{B3}(\bar{\varphi})\dot{\bar{\varphi}}^3 +\frac{1}{8}\left(9\hat{\alpha}_{B}-4\frac{\hat{\alpha}_{B3,\varphi}(\bar{\varphi})}{H_i}\right)\dot{\bar{\varphi}}^4, \nonumber\\
\left(\frac{H}{H_i}\right)^2 ={} &  \left( \frac{\Lambda(\bar{\varphi})}{3H_i^2}-\frac{5}{8}\hat{\alpha}_B\right)  -\frac{3}{4}\hat{\alpha}_{B3}(\bar{\varphi})\dot{\bar{\varphi}}^2 \label{Eq1_bkgd_BUSR}\\
& +\frac{H}{H_i} \hat{\alpha}_{B3}(\bar{\varphi})\dot{\bar{\varphi}}^3 +\frac{1}{8}\left(3\hat{\alpha}_{B}-\frac{4\hat{\alpha}_{B3,\varphi}(\bar{\varphi})}{3H_i}\right)\dot{\bar{\varphi}}^4, \nonumber\\
\left(\frac{H}{H_i}\right)^2 \epsilon  ={} & \frac{3}{4}\dot{\bar{\varphi}}^2[\dot{\bar{\varphi}}^2\hat{\alpha}_{B}-3\hat{\alpha}_{B3}(\bar{\varphi})] -\frac{1}{2}\frac{\hat{\alpha}_{B3,\varphi}(\bar{\varphi})}{H_i}\dot{\bar{\varphi}}^4
\nonumber
\\
&+ \frac{1}{2} \frac{H}{H_i} \hat{\alpha}_{B3}(\bar{\varphi}) \dot{\bar{\varphi}}^3  \left(3-\frac{H_i}{H}\frac{\ddot{\bar{\varphi}}}{H_i\dot{\bar{\varphi}}}\right). \label{Eq2_bkgd_BUSR} 
\end{align}

We will start by considering a particular phase of this model, where $\hat{\alpha}_{B3}=\hat{\alpha}_B$ and $\Lambda$ are both constant. In this case, it is useful to write explicitly the equation of motion for the scalar field as well, in order to illustrate the behaviour of the model. Combining eq.~(\ref{Eq1_bkgd_BUSR})-(\ref{Eq2_bkgd_BUSR}) we obtain:
\begin{equation}\label{Eq3_bkgd_BUSR}
\frac{\ddot{\bar{\varphi}}}{H_i\dot{\bar{\varphi}}}=\frac{3\left(2(H/H_i)\dot{\bar{\varphi}}+\dot{\bar{\varphi}}^2-3\right)\left(-2(H/H_i)+\hat{\alpha}_B\dot{\bar{\varphi}}^3\right)}
{-6+8 (H/H_i)\dot{\bar{\varphi}}+6\dot{\bar{\varphi}}^2+2\hat{\alpha}_B\dot{\bar{\varphi}}^4}.
\end{equation}
We can also solve eq.~(\ref{Eq1_bkgd_BUSR}) to get two solutions of Hubble parameter $H_\pm$:
\begin{equation}\label{Eq_H}
\frac{H_\pm}{H_i} = \frac{\dot{\bar{\varphi}}^3\hat{\alpha}_B}{2} \pm \sqrt{\frac{\dot{\bar{\varphi}}^6\hat{\alpha}_B^2}{4}+\frac{\Lambda}{3H_i^2}+\frac{\hat{\alpha}_B}{8}\left(3\dot{\bar{\varphi}}^4-6\dot{\bar{\varphi}}^2-5\right)}.
\end{equation}
If the sum of the last two terms inside the square root is positive, $H_+$ will always be positive, while $H_-$ will be negative. 
A negative $H$ corresponds to a contracting universe which we are not interested in. Also note that the solution $H_+$ is no less than $H_-$. If $H_+$ is negative then so is $H_-$, so in the following we will focus on the $H_+$ solution.

In the phase that we are interested in, we see that the equations of motion are independent of $\bar{\varphi}$ itself, and hence only $\dot{\bar{\varphi}}$ determines the evolution of the system. 
We find that $\dot{\bar{\varphi}}=0$ is always an attractor solution as we can see from eq.~(\ref{Eq3_bkgd_BUSR}), but it is trivial because the scalar field will stop evolving in time once it hits this solution.
Additionally, if we choose $\Lambda=3H_i^2$, we find another attractor solution $\dot{\bar{\varphi}}=1$, which can be checked by substituting $\dot{\bar{\varphi}}=1$ into eq.~(\ref{Eq3_bkgd_BUSR}). In this case, for $\hat{\alpha}_B<2$, $H_+=H_i$ and the solution gives a de Sitter background with $\ddot{\bar{\varphi}}=0$, so that the scalar field velocity stays constant if initially set to $\dot{\bar{\varphi}}_I=1$. 

For a more concrete example, we choose $\hat{\alpha}_B=1$ and we find three attractor solutions $\dot{\bar{\varphi}}=1,0,-\sqrt{6}$. Note that the number and values of real solutions depend on the choice of $\hat{\alpha}_B$. There are also two repellers $\dot{\bar{\varphi}}\approx 0.677,-0.898$. For this parameter choice, the three attractors are shown in Fig.~\ref{Fig:BUSR_G3Step_PhaseSpace}, where we see that solutions with $\dot{\bar{\varphi}}=1$ evolve towards the right with increasing field values, whereas solutions with $\dot{\bar{\varphi}}=-\sqrt{6}$ evolve towards the left with decreasing field values. 

In what follows, we focus on the $\dot{\bar{\varphi}}=1$ attractor and the $H=H_i$ branch. In this case, the evolution is such that $\epsilon\propto a^{-3}$, for any value of $\hat{\alpha}_B$. This behaviour in $\epsilon$ is similar to that of USR inflation but in this case it relies on the presence of the braiding parameter, and thus we dub this model braiding-ultra-slow-roll (BUSR) inflation.

Next, we analyze the evolution of linear perturbations outside the horizon. This model always has $\alpha_M=\alpha_T=0$, and on the de Sitter attractor we additionally find $\alpha_K=6\hat{\alpha}_B$ and $\alpha_B= \hat{\alpha}_B$. Therefore, we solve the Mukhanov-Sasaki equation using:  
\begin{equation}
Q_s=\frac{3\hat{\alpha}_B(4+\hat{\alpha}_B)}{(2-\hat{\alpha}_B)^2},\quad 
c_s^2=\frac{1}{3}\frac{(2-\hat{\alpha}_B)}{(4+\hat{\alpha}_B)}.
\end{equation}
Note that in order to avoid ghost and gradient instabilities we will need $Q_s>0$ and $c_s^2>0$ \cite{Bellini:2014fua}, and thus we will impose $0<\hat{\alpha}_B<2$.
Both of these coefficients are constants and thus $z\propto a$, and the solution for $u$ will have the same form as eq.~(\ref{FullSolu}), simply generalizing $k$ to $c_s k$.
We then obtain the unitary curvature as $\zeta_u=(u/a)(2Q_s)^{-1/2}$, and see that it will be conserved outside the sound horizon, that is, for $c_sx\ll 1$:
\begin{align}
&\zeta_u \approx \frac{1}{\sqrt{Q_s}}\frac{i H_i}{2(c_sk)^{3/2}} \left[ 1+\frac{1}{2}(c_sx)^2+ \mathcal{O}(x^3)\right],\label{BUSR_Zetau_SuperH}\\
&\zeta_u'\approx -\zeta_u\left[(c_sx)^2+ \mathcal{O}(x^3)\right].\label{BUSR_ZPrime_SuperH}
\end{align}
We emphasize that since this model has a finite braiding parameter $\alpha_B$ and $\epsilon \propto a^{-3}$, then we see from eq.~(\ref{CUcurvature}) that the unitary curvature will grow as $\zeta_c \propto \zeta_u'/\epsilon \propto a$.
Therefore, this inflationary model offers an extreme case where the difference between $\zeta_u$ and $\zeta_c$ can be arbitrarily large. 

Next, we analyze the evolution of $\zeta_s$. From eq.~(\ref{ZetaUS_relation}) we know that in the limit of de Sitter BUSR, where $\epsilon\rightarrow 0$, we have that:
\begin{equation}
    \zeta_s'\approx -\Gamma \zeta_u',
\end{equation}
and therefore $\zeta_s\approx -\Gamma\zeta_u(x=0) [1+\mathcal{O}(\epsilon/\Gamma)]$ (up to some boundary terms).  We thus conclude that for super-horizon modes $\zeta_s' \propto \Gamma (c_sx)^2$ and $\zeta_s$ will be conserved outside the sound horizon. Note that, as found in the previous section, if $\Gamma\lesssim 1$ then $\zeta_s$ will freeze out at the same time or before $\zeta_u$. 

Finally, we calculate the dimensionless power spectrum for modes that have left the horizon already. Since $\zeta_c\propto a$, then the power spectrum of $\zeta_c$ at horizon crossing does not provide meaningful information. Instead, we use the power spectrum of $\zeta_u$ at horizon crossing. From eq.~(\ref{BUSR_Zetau_SuperH}) we find that:
\begin{equation}
    \mathcal{P}_{\zeta_u}(k)= \frac{k^3}{2\pi^2}|\zeta_u|^2= \frac{1}{8\pi^2}\frac{H_i^2}{Q_sc_s^3},
\end{equation}
which is perfectly scale invariant. We can rewrite this in terms of $\hat{\alpha}_B$ as:
\begin{equation}
    \mathcal{P}_{\zeta_u}=  \frac{H_i^2\sqrt{3(2-\hat{\alpha}_B)(4+\hat{\alpha}_B)}}{8\pi^2\hat{\alpha}_B},\label{BUSR_PS}
\end{equation}
and therefore for $\hat{\alpha}_B\approx 1$ we have that $H_i^2$ determines the amplitude of this power spectrum. For an amplitude of order $\mathcal{P}_{\zeta_u}\sim 10^{-9}$ as constrained by temperature anisotropies of the CMB, we would need $H_i\sim 10^{-4}$ (corresponding to an energy scale of $\sqrt{H_i}\sim 10^{16}$GeV). Lower or higher energy scales can be achieved by adjusting the value of $\hat{\alpha}_B$. For instance, for
$H_i\sim 10^{-6}$ (energy scale $10^{15}$GeV) then we must have $\hat{\alpha}_B\sim 10^{-5}$. 
We also note that from the general equations of motion (\ref{Eq1_bkgd_BUSR})-(\ref{Eq2_bkgd_BUSR}) we see that we can always rescale $H_i$ to get an appropriate value for the dimensionless power spectrum $\mathcal{P}_{\zeta_u}(k)$, and the evolution of the system in terms of $e$-folds does not change for the same initial condition $\dot{\bar{\varphi}}=1$, as long as we also rescale $\hat{\alpha}_{B3,\varphi}$ appropriately.

\section{Observables}\label{sec:observables}
In this section, we discuss the consequences of the previous results on the inflationary primordial power spectrum and bispectrum.

\subsection{Separate Universe and $\delta N$ Formalism}\label{sec:shear}
A common technique to calculate the inflationary bispectrum of spatial curvature in a given hypersurface and obtain consistency relations for single-field inflationary models (although it can be extended to more general cases with multiple fields \cite{Lee:2005bb, Matsuda:2009kp,PhysRevD.85.103505} or anisotropic backgrounds \cite{Abolhasani:2013zya, Talebian-Ashkezari:2016llx}) is the $\delta N$ formalism \cite{Sasaki:1995aw, Wands:2000dp, Sugiyama:2012tj,Domenech:2016zxn,Abolhasani:2018gyz,doi:10.1142/10953}. In this formalism, super-horizon perturbations can be reabsorbed into the background equations such that the total perturbed universe still looks homogeneous and isotropic in a local Hubble-sized patch. In this case, the spatial curvature can be calculated as a change in $e$-folds between the effective local and the background universe. 
Although the usage of the $\delta N$ formalism is typically associated with the concept of separate universe, we show that, in some models, violations of separate universe in the matter continuity equation may be present. In this case, as long as the local Hubble rate approximates to that of a Friedmann universe, then the $\delta N$ formula can still be used to accurately compute the power spectrum or bispectrum for appropriate gauge-invariant quantities but subtleties can arise when calculating $\delta N$ to a uniform-density surface.

Formally, we can always write the difference in spatial curvature between some initial and final time slices as \cite{Sasaki:1995aw}:
\begin{equation}\label{InitialDeltaN}
    \Psi(t)-\Psi(t_i)= -\delta N(t;t_i)-\frac{1}{3}\Sigma_I(t;t_i), 
\end{equation}
where we have defined $\delta N$ such that:
\begin{align}
\delta N(t;t_i)&\equiv N_W(t_i,t)-N(t_i,t), \\
N_W(t_i,t) &\equiv \int_{t_i}^t H_W(1+\Phi)dt',\\
N(t_i,t)&\equiv  \int_{t_i}^t Hdt',
\end{align}
where $H_W$ is the total window-averaged Hubble expansion that includes the background and linear perturbations, whereas $H$ is the background Hubble expansion. Similarly, we have defined $\Sigma_I$ as:
\begin{equation}
    \Sigma_I(t,t_i)\equiv \int_{t_i}^t dt'\; H\Sigma,
\end{equation}
where $\Sigma$ is the effective shear of the perturbed expansion of the universe, defined in Section \ref{sec:SeparateUniverse}. 

Eq.~(\ref{InitialDeltaN}) can be used to easily calculate the spatial curvature in terms of background quantities, and hence without solving the perturbed equations of motion. This can be done by first considering an initial time $t_i$ that corresponds to a spatially-flat hypersurface and thus by definition $\Psi(t_i)=0$, whereas the final time can correspond to a more general hypersurface $A$ of interest (e.g.~unitary, comoving, or uniform-density). In addition, we will assume that the shear is negligible in these slices so that we simply obtain that:
\begin{equation}\label{DeltaN}
\zeta_A(t)=-\delta N(t;t_i).
\end{equation}
In most single-field models studied in the literature, the shear in spatially-flat and other slices decays in time and can indeed be neglected. Finally, when spatially-flat observers see an effective local (i.e.~in Hubble-sized patches) homogeneous and isotropic background, it means that super-horizon perturbations can be reabsorbed in such a way that the full perturbed equations take the same form as FRW equations of motion in spatially-flat gauge (this is known as the spatial-gradient expansion where the leading order perturbations in $x \ll 1$ follow FRW-like equations of motion). In this case, super-horizon perturbations can be treated as a homogeneous perturbation to a fiducial background universe on each Hubble horizon scale, leading to the intuitive idea of separate universe where each Hubble-sized patch evolves as an independent effective FRW universe. Computationally, 
$N_W$ can be calculated using just the background equations but changing the initial conditions to reflect the presence of long-wavelength perturbations. Note that these initial conditions do not necessarily result in the same solutions as the global background since they can span a more general class of
homogeneous and isotropic background models.
Specifically, for second-order derivative single-field theories, the spatial curvature can be expressed as:
\begin{equation}
\zeta_A(t)= -N(\varphi_i,\dot{\varphi}_i;t_A)+N(\bar{\varphi}_i,\dot{\bar{\varphi}}_i;t_A), \label{deltaNFinal}
\end{equation}
where $\bar{\varphi}_i$ and $\dot{\bar{\varphi}}_i$ denote the initial conditions for the background scalar field and its derivative, whereas $\varphi_i$ and $\dot{\varphi}_i$ denote the initial conditions for the total scalar field including perturbations and its derivative in spatially-flat gauge. Furthermore, $N$ denotes the number of $e$-folds between the initial and final time according to the Friedmann equation, and both $e$-folds are calculated to a final time $t_A$ such that one reaches the desired hypersurface (e.g.~for obtaining the uniform-density curvature, $t_A$ is chosen such that both $N$ quantities reach the same desired value of the energy density at that time). 

Eq.~(\ref{deltaNFinal}) is known as the $\delta N$ formula which can be used at all orders in perturbations and allow us to calculate the non-linear evolution of $\zeta_A$ by knowing only the background evolution of the theory. In particular, it can be used to obtain $\zeta_A$ at quadratic order for small perturbations. For instance, for models where $\delta \dot{\varphi} $ is negligible, we can Taylor expand the formula and obtain:
\begin{equation}\label{eq:DeltaNTaylor}
\zeta_A(t)= -N_{\varphi}\delta \varphi_i - \frac{1}{2}N_{\varphi\varphi}\delta \varphi^2_i  - \mathcal{O}(\delta\varphi^3),
\end{equation}
which can then be used to calculate e.g.~the squeezed bispectrum for the spatial curvature on comoving slices $\zeta_c$ \cite{Abolhasani:2018gyz}. 

Overall, the validity of the $\delta N$ formula relies on neglecting the shear as well as having the ability to express the equations of motion for perturbations in an FRW-like form so that $N_W$ can be calculated using the background equations of motion. 

Next, we use the USR model to illustrate how the $\delta N$ formula can still be used to calculate the spatial curvature in some slices even though the separate universe condition is technically broken. Nevertheless, the $\delta N$ formula will indeed break for other choices of slices.

Let us suppose we are interested in calculating the spatial curvature in unitary slices, that is, $\zeta_u$. First, we check the conditions of negligible shear. In USR, we find that in spatially-flat gauge at linear order $|\Sigma_{f}| \sim \epsilon\zeta_u$ for super-horizon modes, and since $\epsilon\propto a^{-6}$ we conclude that indeed the shear is negligible in the sense that $|\Sigma_{f}/\zeta_u|\ll 1$. 
Similarly, we also find $|\Sigma_{u}/\zeta_u|\sim x^2\ll 1$. This shows that the first conditions for the validity of separate universe are indeed satisfied for USR. We also mention that in the case of uniform-density curvature we obtain similar results: $|\Sigma_{\rho}/\zeta_\rho|\sim x^2\ll 1$ 
and $|\Sigma_{f}/\zeta_\rho|\sim \epsilon x^{-2}\propto a^{-4}$.

On the other hand, we note that the condition of separate universe defined in synchronous gauge is closely related to the ability of reabsorbing the super-horizon perturbations into the background in spatially-flat gauge. In Section \ref{sec:usr} we found that USR does satisfy the condition  $|\dot{\zeta}_s/(H\zeta_s)|\ll 1$ but the local continuity equation (\ref{FullContinuity}) does not take an FRW-like form as the source term is not negligible. Indeed, it is possible to check that the leading order contribution of the terms in the RHS of eq.~(\ref{SUContinuity}) for super-horizon modes cancel out, hence breaking this condition. In particular, we obtain that the source term evolves as $x^2\zeta_s'\propto a^{-5}$ and the perturbed energy density has the same scaling $\delta \rho_s\propto a^{-5}$, which is different from what we would have expected on the background of this model, as $\rho\propto a^{-6}$. We find the same behaviour in spatially-flat gauge, where there is a technical violation such that in spatially-flat gauge the continuity equation cannot be written in an FRW-like form, whereas  both the (00) equation in (\ref{Full00eq}) and the acceleration equation (\ref{Fullijeq}) can indeed be written like FRW. In this gauge, we again find that $\delta \rho_{f}\propto a^{-5}$. 

As a consequence, in USR we find that the $\delta N $ formula does yield the correct result when calculating $\zeta_u$ (as $N_W\approx N$ with an error falling as $\delta H_f/H\propto \delta \rho_f\propto a^{-5}$) but the incorrect result for the spatial curvature in uniform-density slices $\zeta_{\rho}$. The latter happens because the uniform-density condition for the final surface is misapplied in the separate universe approximation whereas the expansion itself is still FRW to good approximation.

The result for $\zeta_u$ (same as $\zeta_c$ in this model) can be found in \cite{Namjoo:2012aa}.
Here we calculate $\zeta_{\rho}$ to linear order in  perturbations in USR to explicitly illustrate the problem with the $\delta N$ formula. On the one hand, from the result in eq.~(\ref{FullSolu}), it is known from a gauge transformation that $\zeta_\rho=\delta\rho_f/\rho'\approx - H^2x^2 
(\bar{\varphi}'/\rho') (u/a)$ for super-horizon modes.
This expression is always valid for infinitesimal transformations and yields a growing curvature $\zeta_\rho\propto a$, which is the correct result for USR \cite{Romano:2015vxz}. On the other hand, if we use the $\delta N$ formula with the background FRW solutions, we would absorb the spatially flat $\delta \rho_f$ into the background, a constant $\rho$ surface, and evolve to another constant $\rho$ surface leaving $\delta N=0=\zeta_\rho$.   

Finally, we comment on the observable consequence of non-conservation of $\zeta_u$. 
On the one hand, if $\zeta_u$ is indeed conserved outside the horizon, then the resulting squeezed bispectrum will satisfy the standard consistency relation \cite{Maldacena:2002vr,Creminelli:2004yq} given by:
\begin{align}
& \lim_{k_1\rightarrow 0}  
B_{\zeta_u}(k_1,k_2,k_3) =  
\left[ n_s(k_3)-1\right] P_{\zeta_u}(k_1)P_{\zeta_u}(k_3),  
\end{align}
where $k_i$ are three wavenumbers such that $k_2\sim k_3\gg k_1$. Here, $ P_{\zeta_u} =(2\pi^2/k^3){\cal P}_{\zeta_u}$ is the power spectrum with a tilt $(n_s-1)$ defined as:
\begin{equation}
n_s(k)-1=\frac{d\ln {\cal P}_{\zeta_u}(k)}{d\ln k}. 
\end{equation} 
On the other hand, if $\zeta_u$ is not conserved outside the horizon, the squeezed bispectrum will have  a different relation to the power spectrum. In the case of USR, the squeezed bispectrum can be obtained with the $\delta N$ formalism and it is given by \cite{Namjoo:2012aa}:
\begin{align}
& \lim_{k_1\rightarrow 0} B_{\zeta_u}(k_1,k_2,k_3)  = 
6 P_{\zeta_u}(k_1)P_{\zeta_u}(k_3).
\end{align}
However, this relation does not necessarily describe the observable squeezed bispectrum after inflation since the USR phase must end.   Its ending changes the local $e$-folds measured by the observer and the bispectrum accordingly through the $\delta N$ formalism \cite{Cai:2017bxr,Passaglia:2018ixg}.

\subsection{Observable Curvature Power Spectrum}\label{sec:Comoving_vs_Unitary}
In this section, we analyze the observational consequences of inflationary braiding models with a large difference between unitary and comoving curvatures. 

In general, inflationary models provide initial conditions for the seeds of structure formation during early times, which are then used to propagate forward the matter evolution and predict observables such as the Cosmic Microwave Background temperature anisotropies and galaxy distributions. For single-field minimally-coupled inflationary models, these initial conditions are typically obtained by calculating 
$\zeta_u$ at horizon crossing which then determines
$\zeta_c$ during radiation or matter domination outside the horizon. This ultimately becomes the initial 
condition for structure formation.
This translation between unitary and comoving curvatures is straightforwardly done as both curvatures are the same, and both are usually conserved outside the horizon. 

The previously described picture will not be valid anymore in inflationary models with non-minimal couplings if they lead to relevant differences between unitary and comoving curvatures, or if the evolution of these curvatures is not conserved outside the horizon. In such cases, a careful analysis throughout the evolution of inflation until its end must be done to obtain appropriate initial conditions for comoving curvature after inflation. 

As we have previously shown, inflationary models with non-vanishing braiding interactions can lead to a large difference between unitary and comoving curvatures. However, a realistic inflationary model must end and lead to reheating but, as it has been discussed in \cite{Ohashi:2012wf,Lopez:2019wmr}, braiding interactions could potentially spoil the reheating process. For this reason, we will consider BUSR described in Section \ref{sec:busr} and analyze the behaviour of unitary curvature in a setting where braiding vanishes before the end of inflation.
In this scenario, we have that, by construction, unitary and comoving curvatures will coincide at the end of inflation, but we will show that the time evolution of the braiding parameter may lead to a non-trivial evolution of unitary curvature outside the horizon, and therefore its value at horizon crossing may not set the appropriate initial conditions for $\zeta_c$ after inflation. Additionally, we will allow for a time evolution of the potential interactions of BUSR, in order to allow the scalar field to roll down the potential and end the inflationary period.

We start by considering the full BUSR model given by the Horndeski functions in eq.~(\ref{BUSR_G2_Real}). We will make $\hat{\alpha}_{B3}$ evolve in time towards zero, while keeping $\hat{\alpha}_B$ constant. In this case, we keep the kinetic term $X^2$ in $G_2$ the same as before, and change the linear $X$ term together with $G_3$. This is to ensure stability of the model throughout its evolution. We will also allow $\Lambda$ evolve in time, in order to allow for a potential interaction. In this case, BUSR will evolve towards a model with a vanishing braiding parameter as well as a potential-dominated scalar field, as it is considered in common inflationary models. 

In particular, we will consider $\hat{\alpha}_{B3}$ to be a step-like function, starting at $\hat{\alpha}_{B3}=\hat{\alpha}_{B}$ (as in the model of Section \ref{sec:busr}) and ending at zero. For concreteness, we model $\hat{\alpha}_{B3}$ as:
\begin{equation}
\hat{\alpha}_{B3}(\varphi)=\frac{\hat{\alpha}_{B}}{2}\left[1+ \tanh\left(\frac{\varphi_0-\varphi}{d_0}\right)\right],
\end{equation}
where $\varphi_0$ represents the field value where the braiding transition happens, and $d_0$ the width of the transition. Generically, we would also introduce a potential interaction such that:
\begin{equation}
\Lambda(\varphi)=V(\varphi) + \frac{15}{8}H_i^2\hat{\alpha}_B,
\end{equation}
where the first term can be any appropriate inflationary potential, and the second term is added to cancel out the cosmological constant-like term in $G_2(\varphi,X)$ in eq.~(\ref{BUSR_G2_Real}). For simplicity, we will model $V(\varphi)$ with the same $\tanh$ function but with a different width and field location, such that before the step we have $\Lambda\approx 3H_i^2$ and after the step $V\approx 0$. Explicitly, we consider: 
\begin{equation}
\frac{\Lambda(\varphi)}{3 H_i^2}=\frac{1}{2}\left(1- \frac{5}{8}\hat{\alpha}_B\right)\left[1+ \tanh\left(\frac{\varphi_1-\varphi}{d_1}\right)\right]+\frac{5}{8}\hat{\alpha}_B ,
\end{equation}
where $\varphi_1$ represents the field value where the transition in $V$ happens and $d_1$ its width. 

Next, we numerically study the evolution of the model starting from BUSR with initial conditions on the attractor $\dot{\bar{\varphi}}_I=1$ and $\bar{\varphi}_I$ chosen appropriately such that $\hat{\alpha}_{B3}\approx \hat{\alpha}_B$. For concreteness, let us fix the parameters of the model to: $H_i=2.1\times 10^{-4}$, $\hat{\alpha}_B=1$, $\varphi_0=4.7\times 10^4$, $d_0=3.3\times 10^3$, $\varphi_1=5.7\times 10^4$, and $d_1=3.8\times 10^3$. In this case, $\varphi_1>\varphi_0$, and thus the braiding interaction will first vanish, and soon afterwards the scalar field will start rolling down the potential until the end of inflation. We make this choice in order to disentangle effects coming from changing the braiding interaction and those from the potential. 
 \begin{figure}[h!]
	\centering
	\includegraphics[scale=0.60]{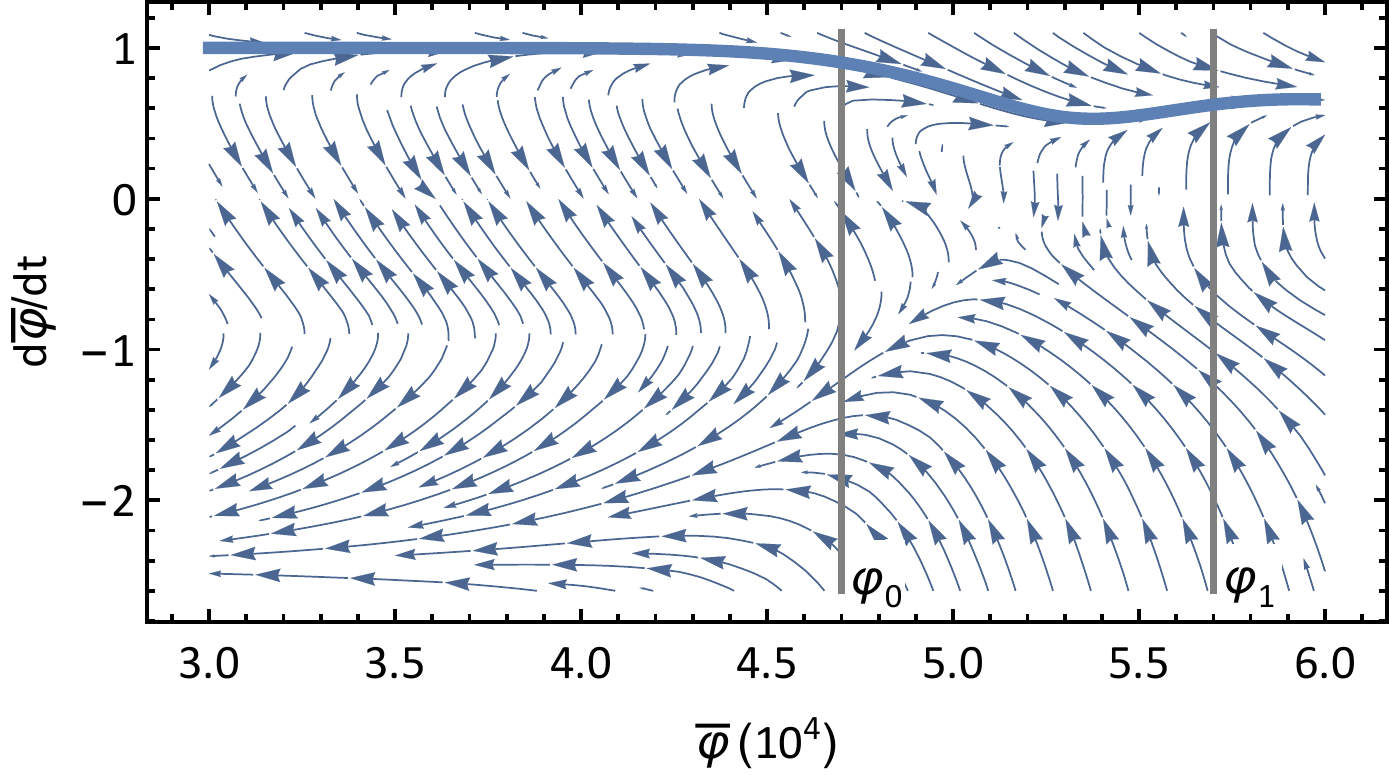}
	\caption{The phase space evolution of the BUSR model through the transition of $G_3(\varphi,X)$. The blue streams show the direction of the time evolution. The blue thick line shows the attractor solution starting with $\dot{\bar{\varphi}}=1$.
	The two vertical gray lines indicate the position of $\varphi_0$ and $\varphi_1$.}
		\label{Fig:BUSR_G3Step_PhaseSpace}
\end{figure}

In Fig.~\ref{Fig:BUSR_G3Step_PhaseSpace} we show the evolution of the scalar field $\bar\varphi$ and its time derivative $\dot{\bar\varphi}$ throughout the transition of $\alpha_{B}$. Starting near  $\dot{\bar\varphi}=1$ on the top left of the plot, we see that flows join the attractor (blue thick curve) as time evolves.
Once  $\bar\varphi\approx \varphi_0$, the transition in $G_3$ occurs, which makes $\dot{\bar\varphi}$ decrease. Then, for $\bar\varphi\approx \varphi_1$ we see that the behaviour changes as the scalar field starts rolling down the potential and its kinetic energy increases.

\begin{figure}[h!]
	\centering
	\includegraphics[scale=0.40]{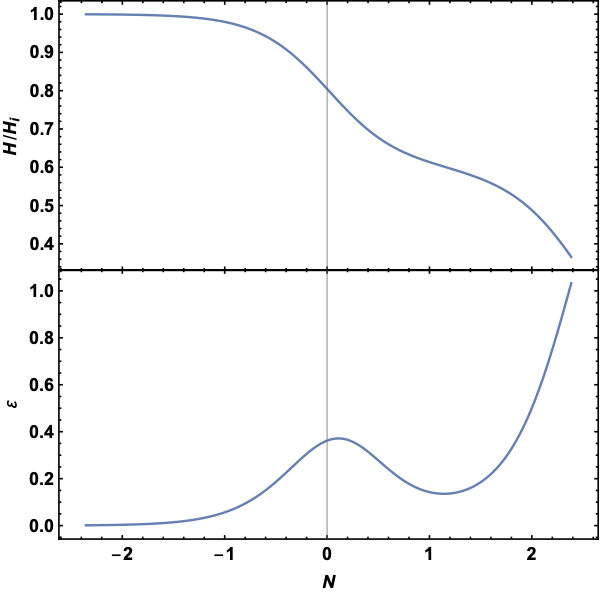}
	\caption{Evolution of $H$ and $\epsilon=-H'/H$ as a function of $e$-folds through the transition of $G_3(\varphi,X)$.}
		\label{Fig:BUSR_G3Step_HEpsilonvsN}
\end{figure}

In Fig.~\ref{Fig:BUSR_G3Step_HEpsilonvsN} we show the evolution of the Hubble rate $H$ and the slow-roll parameter $\epsilon$ as a function of $e$-folds from the braiding transition time (defined as when $\bar\varphi=\varphi_0$). 
In the top panel, we see that $H$ starts near the attractor value $H=H_i$.
Afterwards, the transition in $G_3$ alters the evolution, and $H$ decays towards the value $(H/H_i)=\sqrt{(1-5/8\hat{\alpha}_B)}\approx 0.6$ for $\hat{\alpha}_B=1$. This value of $H$ would be a new de Sitter phase if the potential $V$ was constant, and is obtained from eq.~(\ref{Eq1_bkgd_BUSR}) with $\dot{\bar\varphi}\rightarrow 0$ (i.e.~a potential dominated phase). 
Afterwards, the scalar field starts rolling down the potential and $H$ decays even further towards zero.

In the bottom panel, we see the evolution of $\epsilon$, which starts in a de Sitter phase with $\epsilon\approx 0$. Then, the transition in $G_3$ causes a quick change in $H$ which makes $\epsilon$ grow. Afterwards, $\epsilon$ decays again as the system goes towards the aforementioned would-be de Sitter phase with constant potential, but due to the scalar field rolling down the potential $V$, $\epsilon$ grows again reaching $\epsilon= 1$, marking the end of the accelerated expansion of the universe, and hence the end of inflation.
Note that for the values chosen here, the step in $G_3$ is slow enough that the terms involving $\hat{\alpha}_{B3,\varphi}$ are always negligible in the evolution of $H$ and $\epsilon$. 

Next, we study the evolution of linear cosomological perturbations in this model. Unitary curvature satisfies eq.~(\ref{MukhanovSasaki}). The main coefficients determining the evolution of $\zeta_u$ are thus $c_s^2$ and $Q_s$. In turn, these two quantities depend on the EFT coefficients $\alpha_B$ and $\alpha_K$. Whereas $\alpha_B$ decays to 0 by construction, $\alpha_K$ follows closely the behaviour of the kinetic energy of the scalar field, that is, it decays during the braiding transition, and grows again when the potential dominates the evolution and the scalar field rolls down.

\begin{figure}[h!]
	\centering
	\includegraphics[scale=0.40]{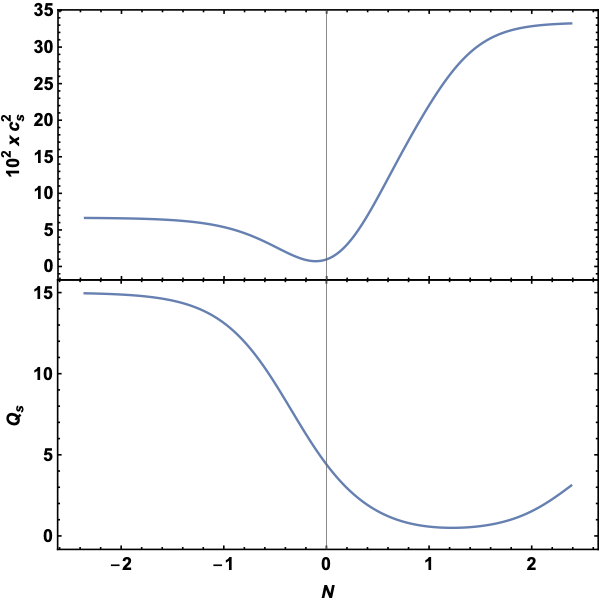}
	\caption{Evolution of $c_s^2$ and $Q_s$ as a function of $e$-folds through the transition of $G_3(\varphi,X)$.}
		\label{Fig:BUSR_G3Step_cs2QsvsN}
\end{figure}

In Fig.~\ref{Fig:BUSR_G3Step_cs2QsvsN}, we explicitly show the evolution of $c_s^2$ and $Q_s$ as a function of $e$-folds when the system undergoes the step in $G_3$. 
In the top panel we see that $c_s^2$ decays during the step of $G_3$. Indeed, the faster the step, the more it decays. Therefore, a fast enough step would lead to a short gradient instability that would potentially spoil the evolution of small-scale perturbations. For this reason, we limit ourselves to cases where the transition in $G_3$ is slow enough to ensure $c_s^2>0$ at all times. After the braiding transition, $c_s^2$ converges towards the value $c_s^2=1/3$ of the final model with $G_2(\varphi,X)=(3/2)H_i^2\hat{\alpha}_BX^2-V(\varphi)$.
In the bottom panel, we see that $Q_s$ decays during the transition of $G_3$ and then grows again when the potential dominates. Overall, $Q_s$ follows a similar behaviour to the kinetic energy of the scalar field.  

Next, we analyze the behaviour of $\zeta_u$ in more detail. From eq.~(\ref{MukhanovSasaki}), the equation of motion for $\zeta_u$ is given by:
\begin{equation}
   \frac{ \left(Ha^3Q_s\zeta_u'\right)'}{Ha^3Q_s}+(c_sx)^2\zeta_u=0,
\end{equation}
and thus from here we generically have that
\begin{equation}\label{ZetaUPrimeIntegral}
    \zeta_u'= -\frac{1}{Ha^3Q_s}\left[  \int d\ln a\,(Ha^3Q_s) (c_sx^2)\zeta_u+\text{const.}\right],
\end{equation}
where we have added an arbitrary integration constant. While this relation is exact, 
it only implicitly determines $\zeta_u$. However, as we shall see next, this expression is useful to analyze $\zeta_u$ in the superhorizon limit. Here, we see that the evolution 
depends crucially on the behavior of $(Ha^3Q_s)$. 

\begin{figure}[h!]
	\centering
	\includegraphics[scale=0.39]{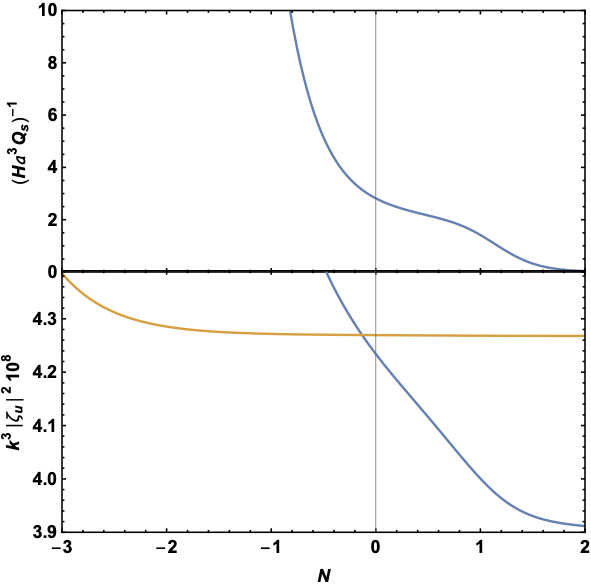}
	\caption{Evolution of $1/(Ha^3Q_s)$ and $k^3|\zeta_u|^2$ for a mode with $x_t\approx 0.6$ (blue line) and a mode with $x_t\approx 0.04$ (yellow line) as a function of $e$-folds through the transition of $G_3(\varphi,X)$.
	}
		\label{Fig:BUSR_G3Step_a3Qs}
\end{figure}

Fig.~\ref{Fig:BUSR_G3Step_a3Qs} shows the time evolution of $(Ha^3Q_s)^{-1}$ (with an arbitrary normalization) and the evolution of $k^3|\zeta_u|^2$ for two modes with $x_t=k/(a_tH_t)$ taking values $x_t\approx 0.6$ and $x_t\approx 0.04$, where $a$ and $H$ are evaluated at the time of the braiding transition (corresponding to $N=0$). In the top panel, we see that $(Ha^3Q_s)^{-1}$ decays monotonically for the value of $d_0$ chosen here, although the decay rate decreases during the braiding transition.
If the transition in $G_3$ is faster, that is for smaller $d_0$, then $(Ha^3Q_s)^{-1}$ may even grow temporarily during the transition. In the bottom panel, we show the evolution of two modes that have already left the horizon before the braiding transition. Whereas very long modes will have a power spectrum at the end of inflation with a value given by eq.~(\ref{BUSR_PS}) (as shown in the yellow line for a mode with $x_t\approx 0.04$), other modes will keep decaying due to the braiding transition (as shown in the blue line for a mode with $x_t\approx 0.6$). This decay will ultimately cause a suppression on the amplitude of the power spectrum on short scales. In the figure, we show the blue mode that leaves the horizon less than 1 $e$-fold before the braiding transition, and its amplitude $k^3|\zeta_u|^2$ at the end of inflation is about $10\%$ smaller than that for modes that leave the horizon well before the transition.

We can understand the behaviour of $\zeta_u$ found in these numerical solutions by using eq.~(\ref{ZetaUPrimeIntegral}). On the one hand, we see that during the de Sitter BUSR phase, $Q_s$, $c_s$ and $H$ are nearly constants, and for modes outside the horizon the integrand of eq.~(\ref{ZetaUPrimeIntegral}) scales as $a$ and, as a consequence, this integral will be dominant over the integration constant. Thus, we find that $\zeta_u'\approx -(c_sx^2)\zeta_u$, in agreement with our previous calculation in eq.~(\ref{BUSR_ZPrime_SuperH}). On the other hand, during the braiding transition, we see that $(Ha^3Q_s)$ becomes roughly constant (this is model dependent as for a given parameter choice $(Ha^3Q_s)$ could be decaying, constant or growing), and thus the behaviour of the integrand will be such that it decays as $a^{-2}$ or even faster. In this case, the total
term inside the square brackets in eq.~(\ref{ZetaUPrimeIntegral}) goes to a constant and we will get that $\zeta_u'\sim $const. We can then match these two behaviours at the time $t_*$ where the braiding transition starts (for our figures, $t_*$ would be between $N=-2$ and $N=-1$), and obtain that during the transition, the behaviour of $\zeta_u$ outside the horizon will be roughly given by $\zeta_u(N)= \zeta_{u*}[1-(c_{s*}x_*)^2(N-N_*)]$. This result shows that for modes that leave the horizon well before the transition (i.e.~$x_*\ll 1$) then the decay rate of $\zeta_u$ during the transition  will be very small. Similarly, for shorter modes that leave the horizon near the transition (e.g.~$x_*\sim 1$), then the decay rate will be much larger. This agrees with the qualitative behaviour shown in Fig.~$\ref{Fig:BUSR_G3Step_a3Qs}$ but because  $(Ha^3Q_s)$ is not exactly constant, the $x_*$ dependence on the slope of $\zeta_u(N)$ is closer to $x_*^{3/2}$ than to $x^2$.

\begin{figure}[h!]
	\centering
	\includegraphics[scale=0.40]{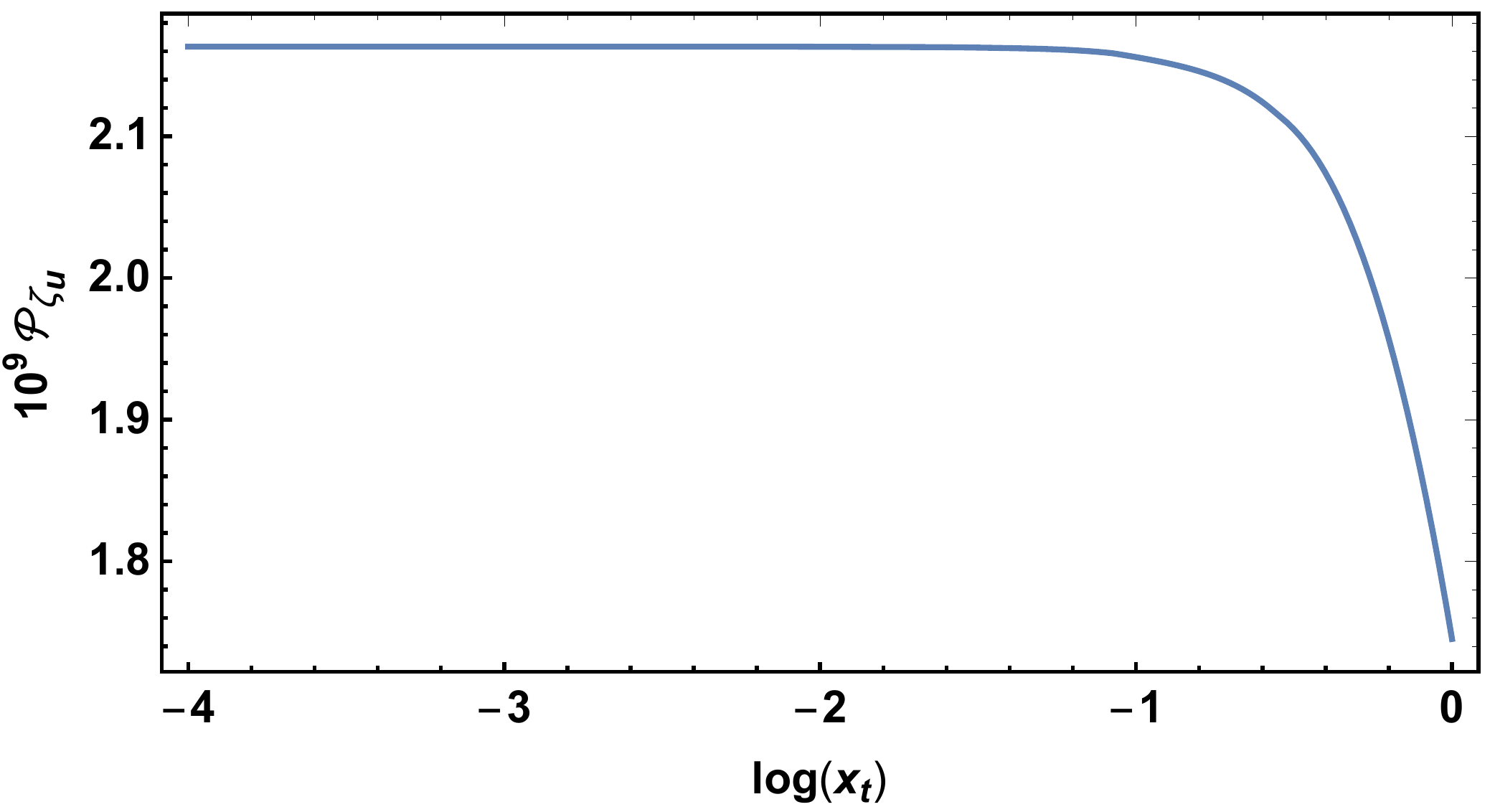}
	\caption{Dimensionless power spectrum ${\cal P}_{\zeta_u}$ at the end of inflation as a function of $x_t=k/(a_t H_t)$.  $x_t<1$ represent modes that were outside the horizon at the time of the transition of $G_3(\varphi,X)$.}
		\label{Fig:BUSR_G3Step_PS}
\end{figure}

Finally, we explicitly calculate the power spectrum at the end of inflation from these modefunctions. Fig.~\ref{Fig:BUSR_G3Step_PS} shows $\mathcal{P}_{\zeta_u}$ as function of $x_t \propto k$.
Here we see that the power spectrum is scale-independent for all modes that left the horizon much earlier than the transition of $G_3$, that is, for $x_t\lesssim 10^{-1}$. However, for modes that leave the horizon right before the braiding transition, there is a suppression of the power spectrum that goes up to $15\%$. From this figure, we find that the decay of $\mathcal{P}_{\zeta_u}$ goes roughly as $x_t^{3/2}$, consistent with the modefunction evolution described above. Note however that this scaling depends on the choice of $\hat{\alpha}_B$ and the width of the braiding transition. Faster steps will lead to a faster decay of $\mathcal{P}_{\zeta_u}$. We therefore conclude that most modes that left the sound horizon in the BUSR phase would indeed have the same primordial amplitude except for a small, but model dependent, number of $e$-folds in $k$ around the sound horizon at the time of the braiding transition.

\section{Discussion}\label{sec:discussion}
In this paper we analyze the difference between three gauge-invariant quantities typically studied in single-field inflationary models: synchronous curvature $\zeta_s$, unitary curvature $\zeta_u$, and comoving curvature $\zeta_c$. We focus on models with second-order derivative equations of motion, and study the evolution of these three curvatures outside the horizon. Generically, we find that conservation of one of these curvatures does not imply conservation of the other two, unless the given model satisfies certain conditions. Conservation of comoving curvature is the most restrictive one, as it will imply conservation of both unitary and synchronous curvatures. 

In order to explicitly illustrate the difference of these three curvatures, we provide two specific examples of inflationary models. First, we discuss the ultra-slow-roll (USR) model, and show that here unitary and comoving curvatures are not conserved outside the horizon (and they are equal), but synchronous curvature is. Second, we construct an inflationary model dubbed braiding-ultra-slow-roll (BUSR), in which there is a non-trivial coupling  that corresponds to indirect derivative interactions (called braiding) between the scalar field and the metric. In this model, we show that unitary and synchronous curvatures are conserved outside the horizon (although they are different), but comoving curvature is not. 

Finally, we discuss the consequences of having different curvatures conserved. First, we clarify that the concept of a separate universe is strictly associated to the synchronous frame since it requires that freely falling observers cannot distinguish that they live in a long-wavelength inhomogeneity through local measurements.  Whereas it is typically assumed that conservation of $\zeta_s$ implies a separate universe, we show that this is not always the case. The USR model is an example where $\zeta_s$ is conserved outside the horizon, but the continuity equation of matter does not approximate to that of a Friedmann universe, and therefore the separate universe approximation is technically broken. 

In addition, we discuss the fact that separate universe is typically considered to be a necessary requirement for the validity of the $\delta N$ formalism, which is in turn used to calculate the primordial power spectrum and bispectrum of inflationary models. We show that although the USR model does mildly violate separate universe, the $\delta N$ formalism can still be used to accurately obtain observables for unitary curvature (even though this is not conserved), but the formalism will break down when calculating observables for uniform-density curvature $\zeta_\rho$.

Finally, we discuss the consequence of a difference in unitary and comoving curvatures, as unitary curvature at horizon crossing is typically used as initial condition for comoving curvature after inflation. We use the BUSR model and start in a phase where $\zeta_u\not=\zeta_c$ due to braiding interactions, but allow the model to evolve towards a phase where braiding eventually goes away. This process allows inflation to end and reheating to start with $\zeta_c=\zeta_u$, reflecting a strong evolution of $\zeta_c$ both after horizon crossing and at the transition to vanishing braiding interaction. The transition can also introduce evolution of the unitary curvature outside the horizon, and thus, at least for some $k$ modes, it would be incorrect to identify $\zeta_u$ at horizon crossing as the initial condition of $\zeta_c$ after inflation. In the particular model considered here, this effect leads to a suppression of power that is confined to scales that are no larger than a few $e$-folds compared to the (sound) horizon at the time of the braiding transition, whereas larger modes that froze out  well before the transition would be unaffected. 

These examples help clarify the concepts of conservation of curvature, separate universe, $\delta N$ and gauge transformations that are often conflated in the literature.

%\section*{Acknowledgments}
%\vspace{-0.2in}
\acknowledgements
We are grateful to Samuel Passaglia for useful comments and discussions. ML was supported by the Kavli Institute for Cosmological Physics at the University of Chicago through an endowment from the Kavli Foundation and its founder Fred Kavli. MXL and WH are supported by the U.S.~Dept.~of  Energy contract DE-FG02-13ER41958 and the Simons Foundation.

\vfill
\bibliography{RefModifiedGravity}

\begin{thebibliography}{39}%
\makeatletter
\providecommand \@ifxundefined [1]{%
 \@ifx{#1\undefined}
}%
\providecommand \@ifnum [1]{%
 \ifnum #1\expandafter \@firstoftwo
 \else \expandafter \@secondoftwo
 \fi
}%
\providecommand \@ifx [1]{%
 \ifx #1\expandafter \@firstoftwo
 \else \expandafter \@secondoftwo
 \fi
}%
\providecommand \natexlab [1]{#1}%
\providecommand \enquote  [1]{``#1''}%
\providecommand \bibnamefont  [1]{#1}%
\providecommand \bibfnamefont [1]{#1}%
\providecommand \citenamefont [1]{#1}%
\providecommand \href@noop [0]{\@secondoftwo}%
\providecommand \href [0]{\begingroup \@sanitize@url \@href}%
\providecommand \@href[1]{\@@startlink{#1}\@@href}%
\providecommand \@@href[1]{\endgroup#1\@@endlink}%
\providecommand \@sanitize@url [0]{\catcode `\\12\catcode `\$12\catcode
  `\&12\catcode `\#12\catcode `\^12\catcode `\_12\catcode `\%12\relax}%
\providecommand \@@startlink[1]{}%
\providecommand \@@endlink[0]{}%
\providecommand \url  [0]{\begingroup\@sanitize@url \@url }%
\providecommand \@url [1]{\endgroup\@href {#1}{\urlprefix }}%
\providecommand \urlprefix  [0]{URL }%
\providecommand \Eprint [0]{\href }%
\providecommand \doibase [0]{http://dx.doi.org/}%
\providecommand \selectlanguage [0]{\@gobble}%
\providecommand \bibinfo  [0]{\@secondoftwo}%
\providecommand \bibfield  [0]{\@secondoftwo}%
\providecommand \translation [1]{[#1]}%
\providecommand \BibitemOpen [0]{}%
\providecommand \bibitemStop [0]{}%
\providecommand \bibitemNoStop [0]{.\EOS\space}%
\providecommand \EOS [0]{\spacefactor3000\relax}%
\providecommand \BibitemShut  [1]{\csname bibitem#1\endcsname}%
\let\auto@bib@innerbib\@empty
%</preamble>
\bibitem [{\citenamefont {Aghanim}\ \emph {et~al.}(2018)\citenamefont {Aghanim}
  \emph {et~al.}}]{Aghanim:2018eyx}%
  \BibitemOpen
  \bibfield  {author} {\bibinfo {author} {\bibfnamefont {N.}~\bibnamefont
  {Aghanim}} \emph {et~al.} (\bibinfo {collaboration} {Planck}),\ }\href@noop
  {} {\  (\bibinfo {year} {2018})},\ \Eprint {http://arxiv.org/abs/1807.06209}
  {arXiv:1807.06209 [astro-ph.CO]} \BibitemShut {NoStop}%
%%CITATION = ARXIV:1807.06209;%%
\bibitem [{\citenamefont {Akrami}\ \emph {et~al.}(2018)\citenamefont {Akrami}
  \emph {et~al.}}]{Akrami:2018odb}%
  \BibitemOpen
  \bibfield  {author} {\bibinfo {author} {\bibfnamefont {Y.}~\bibnamefont
  {Akrami}} \emph {et~al.} (\bibinfo {collaboration} {Planck}),\ }\href@noop {}
  {\  (\bibinfo {year} {2018})},\ \Eprint {http://arxiv.org/abs/1807.06211}
  {arXiv:1807.06211 [astro-ph.CO]} \BibitemShut {NoStop}%
%%CITATION = ARXIV:1807.06211;%%
\bibitem [{\citenamefont {Cheung}\ \emph {et~al.}(2008)\citenamefont {Cheung},
  \citenamefont {Creminelli}, \citenamefont {Fitzpatrick}, \citenamefont
  {Kaplan},\ and\ \citenamefont {Senatore}}]{Cheung:2007st}%
  \BibitemOpen
  \bibfield  {author} {\bibinfo {author} {\bibfnamefont {C.}~\bibnamefont
  {Cheung}}, \bibinfo {author} {\bibfnamefont {P.}~\bibnamefont {Creminelli}},
  \bibinfo {author} {\bibfnamefont {A.~L.}\ \bibnamefont {Fitzpatrick}},
  \bibinfo {author} {\bibfnamefont {J.}~\bibnamefont {Kaplan}}, \ and\ \bibinfo
  {author} {\bibfnamefont {L.}~\bibnamefont {Senatore}},\ }\href {\doibase
  10.1088/1126-6708/2008/03/014} {\bibfield  {journal} {\bibinfo  {journal}
  {JHEP}\ }\textbf {\bibinfo {volume} {03}},\ \bibinfo {pages} {014} (\bibinfo
  {year} {2008})},\ \Eprint {http://arxiv.org/abs/0709.0293} {arXiv:0709.0293
  [hep-th]} \BibitemShut {NoStop}%
%%CITATION = ARXIV:0709.0293;%%
\bibitem [{\citenamefont {Hu}\ \emph {et~al.}(2016)\citenamefont {Hu},
  \citenamefont {Chiang}, \citenamefont {Li},\ and\ \citenamefont
  {LoVerde}}]{Hu:2016ssz}%
  \BibitemOpen
  \bibfield  {author} {\bibinfo {author} {\bibfnamefont {W.}~\bibnamefont
  {Hu}}, \bibinfo {author} {\bibfnamefont {C.-T.}\ \bibnamefont {Chiang}},
  \bibinfo {author} {\bibfnamefont {Y.}~\bibnamefont {Li}}, \ and\ \bibinfo
  {author} {\bibfnamefont {M.}~\bibnamefont {LoVerde}},\ }\href {\doibase
  10.1103/PhysRevD.94.023002} {\bibfield  {journal} {\bibinfo  {journal} {Phys.
  Rev.}\ }\textbf {\bibinfo {volume} {D94}},\ \bibinfo {pages} {023002}
  (\bibinfo {year} {2016})},\ \Eprint {http://arxiv.org/abs/1605.01412}
  {arXiv:1605.01412 [astro-ph.CO]} \BibitemShut {NoStop}%
%%CITATION = ARXIV:1605.01412;%%
\bibitem [{\citenamefont {Sasaki}\ and\ \citenamefont
  {Stewart}(1996)}]{Sasaki:1995aw}%
  \BibitemOpen
  \bibfield  {author} {\bibinfo {author} {\bibfnamefont {M.}~\bibnamefont
  {Sasaki}}\ and\ \bibinfo {author} {\bibfnamefont {E.~D.}\ \bibnamefont
  {Stewart}},\ }\href {\doibase 10.1143/PTP.95.71} {\bibfield  {journal}
  {\bibinfo  {journal} {Prog. Theor. Phys.}\ }\textbf {\bibinfo {volume}
  {95}},\ \bibinfo {pages} {71} (\bibinfo {year} {1996})},\ \Eprint
  {http://arxiv.org/abs/astro-ph/9507001} {arXiv:astro-ph/9507001 [astro-ph]}
  \BibitemShut {NoStop}%
%%CITATION = ASTRO-PH/9507001;%%
\bibitem [{\citenamefont {Wands}\ \emph {et~al.}(2000)\citenamefont {Wands},
  \citenamefont {Malik}, \citenamefont {Lyth},\ and\ \citenamefont
  {Liddle}}]{Wands:2000dp}%
  \BibitemOpen
  \bibfield  {author} {\bibinfo {author} {\bibfnamefont {D.}~\bibnamefont
  {Wands}}, \bibinfo {author} {\bibfnamefont {K.~A.}\ \bibnamefont {Malik}},
  \bibinfo {author} {\bibfnamefont {D.~H.}\ \bibnamefont {Lyth}}, \ and\
  \bibinfo {author} {\bibfnamefont {A.~R.}\ \bibnamefont {Liddle}},\ }\href
  {\doibase 10.1103/PhysRevD.62.043527} {\bibfield  {journal} {\bibinfo
  {journal} {Phys. Rev.}\ }\textbf {\bibinfo {volume} {D62}},\ \bibinfo {pages}
  {043527} (\bibinfo {year} {2000})},\ \Eprint
  {http://arxiv.org/abs/astro-ph/0003278} {arXiv:astro-ph/0003278 [astro-ph]}
  \BibitemShut {NoStop}%
%%CITATION = ASTRO-PH/0003278;%%
\bibitem [{\citenamefont {Sugiyama}\ \emph {et~al.}(2013)\citenamefont
  {Sugiyama}, \citenamefont {Komatsu},\ and\ \citenamefont
  {Futamase}}]{Sugiyama:2012tj}%
  \BibitemOpen
  \bibfield  {author} {\bibinfo {author} {\bibfnamefont {N.~S.}\ \bibnamefont
  {Sugiyama}}, \bibinfo {author} {\bibfnamefont {E.}~\bibnamefont {Komatsu}}, \
  and\ \bibinfo {author} {\bibfnamefont {T.}~\bibnamefont {Futamase}},\ }\href
  {\doibase 10.1103/PhysRevD.87.023530} {\bibfield  {journal} {\bibinfo
  {journal} {Phys. Rev.}\ }\textbf {\bibinfo {volume} {D87}},\ \bibinfo {pages}
  {023530} (\bibinfo {year} {2013})},\ \Eprint {http://arxiv.org/abs/1208.1073}
  {arXiv:1208.1073 [gr-qc]} \BibitemShut {NoStop}%
%%CITATION = ARXIV:1208.1073;%%
\bibitem [{\citenamefont {Domenech}\ \emph {et~al.}(2017)\citenamefont
  {Domenech}, \citenamefont {Gong},\ and\ \citenamefont
  {Sasaki}}]{Domenech:2016zxn}%
  \BibitemOpen
  \bibfield  {author} {\bibinfo {author} {\bibfnamefont {G.}~\bibnamefont
  {Domenech}}, \bibinfo {author} {\bibfnamefont {J.-O.}\ \bibnamefont {Gong}},
  \ and\ \bibinfo {author} {\bibfnamefont {M.}~\bibnamefont {Sasaki}},\ }\href
  {\doibase 10.1016/j.physletb.2017.04.014} {\bibfield  {journal} {\bibinfo
  {journal} {Phys. Lett.}\ }\textbf {\bibinfo {volume} {B769}},\ \bibinfo
  {pages} {413} (\bibinfo {year} {2017})},\ \Eprint
  {http://arxiv.org/abs/1606.03343} {arXiv:1606.03343 [astro-ph.CO]}
  \BibitemShut {NoStop}%
%%CITATION = ARXIV:1606.03343;%%
\bibitem [{\citenamefont {Abolhasani}\ and\ \citenamefont
  {Sasaki}(2018)}]{Abolhasani:2018gyz}%
  \BibitemOpen
  \bibfield  {author} {\bibinfo {author} {\bibfnamefont {A.~A.}\ \bibnamefont
  {Abolhasani}}\ and\ \bibinfo {author} {\bibfnamefont {M.}~\bibnamefont
  {Sasaki}},\ }\href {\doibase 10.1088/1475-7516/2018/08/025} {\bibfield
  {journal} {\bibinfo  {journal} {JCAP}\ }\textbf {\bibinfo {volume} {1808}},\
  \bibinfo {pages} {025} (\bibinfo {year} {2018})},\ \Eprint
  {http://arxiv.org/abs/1805.11298} {arXiv:1805.11298 [astro-ph.CO]}
  \BibitemShut {NoStop}%
%%CITATION = ARXIV:1805.11298;%%
\bibitem [{\citenamefont {Abolhasani}\ \emph {et~al.}(2019)\citenamefont
  {Abolhasani}, \citenamefont {Firouzjahi}, \citenamefont {Naruko},\ and\
  \citenamefont {Sasaki}}]{doi:10.1142/10953}%
  \BibitemOpen
  \bibfield  {author} {\bibinfo {author} {\bibfnamefont {A.~A.}\ \bibnamefont
  {Abolhasani}}, \bibinfo {author} {\bibfnamefont {H.}~\bibnamefont
  {Firouzjahi}}, \bibinfo {author} {\bibfnamefont {A.}~\bibnamefont {Naruko}},
  \ and\ \bibinfo {author} {\bibfnamefont {M.}~\bibnamefont {Sasaki}},\ }\href
  {\doibase 10.1142/10953} {\emph {\bibinfo {title} {Delta N Formalism in
  Cosmological Perturbation Theory}}}\ (\bibinfo  {publisher} {WORLD
  SCIENTIFIC},\ \bibinfo {year} {2019})\ \Eprint
  {http://arxiv.org/abs/https://www.worldscientific.com/doi/pdf/10.1142/10953}
  {https://www.worldscientific.com/doi/pdf/10.1142/10953} \BibitemShut
  {NoStop}%
\bibitem [{\citenamefont {Kinney}(2005)}]{Kinney:2005vj}%
  \BibitemOpen
  \bibfield  {author} {\bibinfo {author} {\bibfnamefont {W.~H.}\ \bibnamefont
  {Kinney}},\ }\href {\doibase 10.1103/PhysRevD.72.023515} {\bibfield
  {journal} {\bibinfo  {journal} {Phys. Rev.}\ }\textbf {\bibinfo {volume}
  {D72}},\ \bibinfo {pages} {023515} (\bibinfo {year} {2005})},\ \Eprint
  {http://arxiv.org/abs/gr-qc/0503017} {arXiv:gr-qc/0503017 [gr-qc]}
  \BibitemShut {NoStop}%
%%CITATION = GR-QC/0503017;%%
\bibitem [{\citenamefont {Hu}\ and\ \citenamefont {Joyce}(2017)}]{Hu:2016wfa}%
  \BibitemOpen
  \bibfield  {author} {\bibinfo {author} {\bibfnamefont {W.}~\bibnamefont
  {Hu}}\ and\ \bibinfo {author} {\bibfnamefont {A.}~\bibnamefont {Joyce}},\
  }\href {\doibase 10.1103/PhysRevD.95.043529} {\bibfield  {journal} {\bibinfo
  {journal} {Phys. Rev.}\ }\textbf {\bibinfo {volume} {D95}},\ \bibinfo {pages}
  {043529} (\bibinfo {year} {2017})},\ \Eprint
  {http://arxiv.org/abs/1612.02454} {arXiv:1612.02454 [astro-ph.CO]}
  \BibitemShut {NoStop}%
%%CITATION = ARXIV:1612.02454;%%
\bibitem [{\citenamefont {Lyth}\ \emph {et~al.}(2005)\citenamefont {Lyth},
  \citenamefont {Malik},\ and\ \citenamefont {Sasaki}}]{Lyth:2004gb}%
  \BibitemOpen
  \bibfield  {author} {\bibinfo {author} {\bibfnamefont {D.~H.}\ \bibnamefont
  {Lyth}}, \bibinfo {author} {\bibfnamefont {K.~A.}\ \bibnamefont {Malik}}, \
  and\ \bibinfo {author} {\bibfnamefont {M.}~\bibnamefont {Sasaki}},\ }\href
  {\doibase 10.1088/1475-7516/2005/05/004} {\bibfield  {journal} {\bibinfo
  {journal} {JCAP}\ }\textbf {\bibinfo {volume} {0505}},\ \bibinfo {pages}
  {004} (\bibinfo {year} {2005})},\ \Eprint
  {http://arxiv.org/abs/astro-ph/0411220} {arXiv:astro-ph/0411220 [astro-ph]}
  \BibitemShut {NoStop}%
%%CITATION = ASTRO-PH/0411220;%%
\bibitem [{\citenamefont {Ohashi}\ and\ \citenamefont
  {Tsujikawa}(2012)}]{Ohashi:2012wf}%
  \BibitemOpen
  \bibfield  {author} {\bibinfo {author} {\bibfnamefont {J.}~\bibnamefont
  {Ohashi}}\ and\ \bibinfo {author} {\bibfnamefont {S.}~\bibnamefont
  {Tsujikawa}},\ }\href {\doibase 10.1088/1475-7516/2012/10/035} {\bibfield
  {journal} {\bibinfo  {journal} {JCAP}\ }\textbf {\bibinfo {volume} {1210}},\
  \bibinfo {pages} {035} (\bibinfo {year} {2012})},\ \Eprint
  {http://arxiv.org/abs/1207.4879} {arXiv:1207.4879 [gr-qc]} \BibitemShut
  {NoStop}%
%%CITATION = ARXIV:1207.4879;%%
\bibitem [{\citenamefont {Lopez}\ \emph {et~al.}(2019)\citenamefont {Lopez},
  \citenamefont {Maggiolo}, \citenamefont {Videla}, \citenamefont {Gonzalez},\
  and\ \citenamefont {Panotopoulos}}]{Lopez:2019wmr}%
  \BibitemOpen
  \bibfield  {author} {\bibinfo {author} {\bibfnamefont {M.}~\bibnamefont
  {Lopez}}, \bibinfo {author} {\bibfnamefont {J.}~\bibnamefont {Maggiolo}},
  \bibinfo {author} {\bibfnamefont {N.}~\bibnamefont {Videla}}, \bibinfo
  {author} {\bibfnamefont {P.}~\bibnamefont {Gonzalez}}, \ and\ \bibinfo
  {author} {\bibfnamefont {G.}~\bibnamefont {Panotopoulos}},\ }\href@noop {} {\
   (\bibinfo {year} {2019})},\ \Eprint {http://arxiv.org/abs/1908.03155}
  {arXiv:1908.03155 [gr-qc]} \BibitemShut {NoStop}%
%%CITATION = ARXIV:1908.03155;%%
\bibitem [{\citenamefont {Bardeen}(1980)}]{Bardeen:1980kt}%
  \BibitemOpen
  \bibfield  {author} {\bibinfo {author} {\bibfnamefont {J.~M.}\ \bibnamefont
  {Bardeen}},\ }\href {\doibase 10.1103/PhysRevD.22.1882} {\bibfield  {journal}
  {\bibinfo  {journal} {Phys. Rev.}\ }\textbf {\bibinfo {volume} {D22}},\
  \bibinfo {pages} {1882} (\bibinfo {year} {1980})}\BibitemShut {NoStop}%
%%CITATION = PHRVA,D22,1882;%%
\bibitem [{\citenamefont {Horndeski}(1974)}]{Horndeski:1974wa}%
  \BibitemOpen
  \bibfield  {author} {\bibinfo {author} {\bibfnamefont {G.~W.}\ \bibnamefont
  {Horndeski}},\ }\href {\doibase 10.1007/BF01807638} {\bibfield  {journal}
  {\bibinfo  {journal} {Int. J. Theor. Phys.}\ }\textbf {\bibinfo {volume}
  {10}},\ \bibinfo {pages} {363} (\bibinfo {year} {1974})}\BibitemShut
  {NoStop}%
%%CITATION = IJTPB,10,363;%%
\bibitem [{\citenamefont {Deffayet}\ \emph {et~al.}(2011)\citenamefont
  {Deffayet}, \citenamefont {Gao}, \citenamefont {Steer},\ and\ \citenamefont
  {Zahariade}}]{Deffayet:2011gz}%
  \BibitemOpen
  \bibfield  {author} {\bibinfo {author} {\bibfnamefont {C.}~\bibnamefont
  {Deffayet}}, \bibinfo {author} {\bibfnamefont {X.}~\bibnamefont {Gao}},
  \bibinfo {author} {\bibfnamefont {D.~A.}\ \bibnamefont {Steer}}, \ and\
  \bibinfo {author} {\bibfnamefont {G.}~\bibnamefont {Zahariade}},\ }\href
  {\doibase 10.1103/PhysRevD.84.064039} {\bibfield  {journal} {\bibinfo
  {journal} {Phys. Rev.}\ }\textbf {\bibinfo {volume} {D84}},\ \bibinfo {pages}
  {064039} (\bibinfo {year} {2011})},\ \Eprint {http://arxiv.org/abs/1103.3260}
  {arXiv:1103.3260 [hep-th]} \BibitemShut {NoStop}%
%%CITATION = ARXIV:1103.3260;%%
\bibitem [{\citenamefont {Bellini}\ and\ \citenamefont
  {Sawicki}(2014)}]{Bellini:2014fua}%
  \BibitemOpen
  \bibfield  {author} {\bibinfo {author} {\bibfnamefont {E.}~\bibnamefont
  {Bellini}}\ and\ \bibinfo {author} {\bibfnamefont {I.}~\bibnamefont
  {Sawicki}},\ }\href {\doibase 10.1088/1475-7516/2014/07/050} {\bibfield
  {journal} {\bibinfo  {journal} {JCAP}\ }\textbf {\bibinfo {volume} {1407}},\
  \bibinfo {pages} {050} (\bibinfo {year} {2014})},\ \Eprint
  {http://arxiv.org/abs/1404.3713} {arXiv:1404.3713 [astro-ph.CO]} \BibitemShut
  {NoStop}%
%%CITATION = ARXIV:1404.3713;%%
\bibitem [{\citenamefont {Motohashi}\ and\ \citenamefont
  {Hu}(2017)}]{Motohashi:2017gqb}%
  \BibitemOpen
  \bibfield  {author} {\bibinfo {author} {\bibfnamefont {H.}~\bibnamefont
  {Motohashi}}\ and\ \bibinfo {author} {\bibfnamefont {W.}~\bibnamefont {Hu}},\
  }\href {\doibase 10.1103/PhysRevD.96.023502} {\bibfield  {journal} {\bibinfo
  {journal} {Phys. Rev.}\ }\textbf {\bibinfo {volume} {D96}},\ \bibinfo {pages}
  {023502} (\bibinfo {year} {2017})},\ \Eprint
  {http://arxiv.org/abs/1704.01128} {arXiv:1704.01128 [hep-th]} \BibitemShut
  {NoStop}%
%%CITATION = ARXIV:1704.01128;%%
\bibitem [{\citenamefont {Namjoo}\ \emph {et~al.}(2013)\citenamefont {Namjoo},
  \citenamefont {Firouzjahi},\ and\ \citenamefont {Sasaki}}]{Namjoo:2012aa}%
  \BibitemOpen
  \bibfield  {author} {\bibinfo {author} {\bibfnamefont {M.~H.}\ \bibnamefont
  {Namjoo}}, \bibinfo {author} {\bibfnamefont {H.}~\bibnamefont {Firouzjahi}},
  \ and\ \bibinfo {author} {\bibfnamefont {M.}~\bibnamefont {Sasaki}},\ }\href
  {\doibase 10.1209/0295-5075/101/39001} {\bibfield  {journal} {\bibinfo
  {journal} {EPL}\ }\textbf {\bibinfo {volume} {101}},\ \bibinfo {pages}
  {39001} (\bibinfo {year} {2013})},\ \Eprint {http://arxiv.org/abs/1210.3692}
  {arXiv:1210.3692 [astro-ph.CO]} \BibitemShut {NoStop}%
%%CITATION = ARXIV:1210.3692;%%
\bibitem [{\citenamefont {Martin}\ \emph {et~al.}(2013)\citenamefont {Martin},
  \citenamefont {Motohashi},\ and\ \citenamefont {Suyama}}]{Martin:2012pe}%
  \BibitemOpen
  \bibfield  {author} {\bibinfo {author} {\bibfnamefont {J.}~\bibnamefont
  {Martin}}, \bibinfo {author} {\bibfnamefont {H.}~\bibnamefont {Motohashi}}, \
  and\ \bibinfo {author} {\bibfnamefont {T.}~\bibnamefont {Suyama}},\ }\href
  {\doibase 10.1103/PhysRevD.87.023514} {\bibfield  {journal} {\bibinfo
  {journal} {Phys. Rev.}\ }\textbf {\bibinfo {volume} {D87}},\ \bibinfo {pages}
  {023514} (\bibinfo {year} {2013})},\ \Eprint {http://arxiv.org/abs/1211.0083}
  {arXiv:1211.0083 [astro-ph.CO]} \BibitemShut {NoStop}%
%%CITATION = ARXIV:1211.0083;%%
\bibitem [{\citenamefont {Huang}\ and\ \citenamefont
  {Wang}(2013)}]{Huang:2013lda}%
  \BibitemOpen
  \bibfield  {author} {\bibinfo {author} {\bibfnamefont {Q.-G.}\ \bibnamefont
  {Huang}}\ and\ \bibinfo {author} {\bibfnamefont {Y.}~\bibnamefont {Wang}},\
  }\href {\doibase 10.1088/1475-7516/2013/06/035} {\bibfield  {journal}
  {\bibinfo  {journal} {JCAP}\ }\textbf {\bibinfo {volume} {1306}},\ \bibinfo
  {pages} {035} (\bibinfo {year} {2013})},\ \Eprint
  {http://arxiv.org/abs/1303.4526} {arXiv:1303.4526 [hep-th]} \BibitemShut
  {NoStop}%
%%CITATION = ARXIV:1303.4526;%%
\bibitem [{\citenamefont {Mooij}\ and\ \citenamefont
  {Palma}(2015)}]{Mooij:2015yka}%
  \BibitemOpen
  \bibfield  {author} {\bibinfo {author} {\bibfnamefont {S.}~\bibnamefont
  {Mooij}}\ and\ \bibinfo {author} {\bibfnamefont {G.~A.}\ \bibnamefont
  {Palma}},\ }\href {\doibase 10.1088/1475-7516/2015/11/025} {\bibfield
  {journal} {\bibinfo  {journal} {JCAP}\ }\textbf {\bibinfo {volume} {1511}},\
  \bibinfo {pages} {025} (\bibinfo {year} {2015})},\ \Eprint
  {http://arxiv.org/abs/1502.03458} {arXiv:1502.03458 [astro-ph.CO]}
  \BibitemShut {NoStop}%
%%CITATION = ARXIV:1502.03458;%%
\bibitem [{\citenamefont {Romano}\ \emph
  {et~al.}(2016{\natexlab{a}})\citenamefont {Romano}, \citenamefont {Mooij},\
  and\ \citenamefont {Sasaki}}]{Romano:2016gop}%
  \BibitemOpen
  \bibfield  {author} {\bibinfo {author} {\bibfnamefont {A.~E.}\ \bibnamefont
  {Romano}}, \bibinfo {author} {\bibfnamefont {S.}~\bibnamefont {Mooij}}, \
  and\ \bibinfo {author} {\bibfnamefont {M.}~\bibnamefont {Sasaki}},\ }\href
  {\doibase 10.1016/j.physletb.2016.08.025} {\bibfield  {journal} {\bibinfo
  {journal} {Phys. Lett.}\ }\textbf {\bibinfo {volume} {B761}},\ \bibinfo
  {pages} {119} (\bibinfo {year} {2016}{\natexlab{a}})},\ \Eprint
  {http://arxiv.org/abs/1606.04906} {arXiv:1606.04906 [gr-qc]} \BibitemShut
  {NoStop}%
%%CITATION = ARXIV:1606.04906;%%
\bibitem [{\citenamefont {Bravo}\ \emph {et~al.}(2018)\citenamefont {Bravo},
  \citenamefont {Mooij}, \citenamefont {Palma},\ and\ \citenamefont
  {Pradenas}}]{Bravo:2017wyw}%
  \BibitemOpen
  \bibfield  {author} {\bibinfo {author} {\bibfnamefont {R.}~\bibnamefont
  {Bravo}}, \bibinfo {author} {\bibfnamefont {S.}~\bibnamefont {Mooij}},
  \bibinfo {author} {\bibfnamefont {G.~A.}\ \bibnamefont {Palma}}, \ and\
  \bibinfo {author} {\bibfnamefont {B.}~\bibnamefont {Pradenas}},\ }\href
  {\doibase 10.1088/1475-7516/2018/05/024} {\bibfield  {journal} {\bibinfo
  {journal} {JCAP}\ }\textbf {\bibinfo {volume} {1805}},\ \bibinfo {pages}
  {024} (\bibinfo {year} {2018})},\ \Eprint {http://arxiv.org/abs/1711.02680}
  {arXiv:1711.02680 [astro-ph.CO]} \BibitemShut {NoStop}%
%%CITATION = ARXIV:1711.02680;%%
\bibitem [{\citenamefont {Germani}\ and\ \citenamefont
  {Prokopec}(2017)}]{Germani:2017bcs}%
  \BibitemOpen
  \bibfield  {author} {\bibinfo {author} {\bibfnamefont {C.}~\bibnamefont
  {Germani}}\ and\ \bibinfo {author} {\bibfnamefont {T.}~\bibnamefont
  {Prokopec}},\ }\href {\doibase 10.1016/j.dark.2017.09.001} {\bibfield
  {journal} {\bibinfo  {journal} {Phys. Dark Univ.}\ }\textbf {\bibinfo
  {volume} {18}},\ \bibinfo {pages} {6} (\bibinfo {year} {2017})},\ \Eprint
  {http://arxiv.org/abs/1706.04226} {arXiv:1706.04226 [astro-ph.CO]}
  \BibitemShut {NoStop}%
%%CITATION = ARXIV:1706.04226;%%
\bibitem [{\citenamefont {Cai}\ \emph {et~al.}(2018)\citenamefont {Cai},
  \citenamefont {Chen}, \citenamefont {Namjoo}, \citenamefont {Sasaki},
  \citenamefont {Wang},\ and\ \citenamefont {Wang}}]{Cai:2017bxr}%
  \BibitemOpen
  \bibfield  {author} {\bibinfo {author} {\bibfnamefont {Y.-F.}\ \bibnamefont
  {Cai}}, \bibinfo {author} {\bibfnamefont {X.}~\bibnamefont {Chen}}, \bibinfo
  {author} {\bibfnamefont {M.~H.}\ \bibnamefont {Namjoo}}, \bibinfo {author}
  {\bibfnamefont {M.}~\bibnamefont {Sasaki}}, \bibinfo {author} {\bibfnamefont
  {D.-G.}\ \bibnamefont {Wang}}, \ and\ \bibinfo {author} {\bibfnamefont
  {Z.}~\bibnamefont {Wang}},\ }\href {\doibase 10.1088/1475-7516/2018/05/012}
  {\bibfield  {journal} {\bibinfo  {journal} {JCAP}\ }\textbf {\bibinfo
  {volume} {1805}},\ \bibinfo {pages} {012} (\bibinfo {year} {2018})},\ \Eprint
  {http://arxiv.org/abs/1712.09998} {arXiv:1712.09998 [astro-ph.CO]}
  \BibitemShut {NoStop}%
%%CITATION = ARXIV:1712.09998;%%
\bibitem [{\citenamefont {Passaglia}\ \emph {et~al.}(2019)\citenamefont
  {Passaglia}, \citenamefont {Hu},\ and\ \citenamefont
  {Motohashi}}]{Passaglia:2018ixg}%
  \BibitemOpen
  \bibfield  {author} {\bibinfo {author} {\bibfnamefont {S.}~\bibnamefont
  {Passaglia}}, \bibinfo {author} {\bibfnamefont {W.}~\bibnamefont {Hu}}, \
  and\ \bibinfo {author} {\bibfnamefont {H.}~\bibnamefont {Motohashi}},\ }\href
  {\doibase 10.1103/PhysRevD.99.043536} {\bibfield  {journal} {\bibinfo
  {journal} {Phys. Rev.}\ }\textbf {\bibinfo {volume} {D99}},\ \bibinfo {pages}
  {043536} (\bibinfo {year} {2019})},\ \Eprint
  {http://arxiv.org/abs/1812.08243} {arXiv:1812.08243 [astro-ph.CO]}
  \BibitemShut {NoStop}%
%%CITATION = ARXIV:1812.08243;%%
\bibitem [{\citenamefont {Kennedy}\ \emph {et~al.}(2017)\citenamefont
  {Kennedy}, \citenamefont {Lombriser},\ and\ \citenamefont
  {Taylor}}]{Kennedy:2017sof}%
  \BibitemOpen
  \bibfield  {author} {\bibinfo {author} {\bibfnamefont {J.}~\bibnamefont
  {Kennedy}}, \bibinfo {author} {\bibfnamefont {L.}~\bibnamefont {Lombriser}},
  \ and\ \bibinfo {author} {\bibfnamefont {A.}~\bibnamefont {Taylor}},\ }\href
  {\doibase 10.1103/PhysRevD.96.084051} {\bibfield  {journal} {\bibinfo
  {journal} {Phys. Rev.}\ }\textbf {\bibinfo {volume} {D96}},\ \bibinfo {pages}
  {084051} (\bibinfo {year} {2017})},\ \Eprint
  {http://arxiv.org/abs/1705.09290} {arXiv:1705.09290 [gr-qc]} \BibitemShut
  {NoStop}%
%%CITATION = ARXIV:1705.09290;%%
\bibitem [{\citenamefont {Kennedy}\ \emph {et~al.}(2018)\citenamefont
  {Kennedy}, \citenamefont {Lombriser},\ and\ \citenamefont
  {Taylor}}]{Kennedy:2018gtx}%
  \BibitemOpen
  \bibfield  {author} {\bibinfo {author} {\bibfnamefont {J.}~\bibnamefont
  {Kennedy}}, \bibinfo {author} {\bibfnamefont {L.}~\bibnamefont {Lombriser}},
  \ and\ \bibinfo {author} {\bibfnamefont {A.}~\bibnamefont {Taylor}},\ }\href
  {\doibase 10.1103/PhysRevD.98.044051} {\bibfield  {journal} {\bibinfo
  {journal} {Phys. Rev.}\ }\textbf {\bibinfo {volume} {D98}},\ \bibinfo {pages}
  {044051} (\bibinfo {year} {2018})},\ \Eprint
  {http://arxiv.org/abs/1804.04582} {arXiv:1804.04582 [astro-ph.CO]}
  \BibitemShut {NoStop}%
%%CITATION = ARXIV:1804.04582;%%
\bibitem [{\citenamefont {Lee}\ \emph {et~al.}(2005)\citenamefont {Lee},
  \citenamefont {Sasaki}, \citenamefont {Stewart}, \citenamefont {Tanaka},\
  and\ \citenamefont {Yokoyama}}]{Lee:2005bb}%
  \BibitemOpen
  \bibfield  {author} {\bibinfo {author} {\bibfnamefont {H.-C.}\ \bibnamefont
  {Lee}}, \bibinfo {author} {\bibfnamefont {M.}~\bibnamefont {Sasaki}},
  \bibinfo {author} {\bibfnamefont {E.~D.}\ \bibnamefont {Stewart}}, \bibinfo
  {author} {\bibfnamefont {T.}~\bibnamefont {Tanaka}}, \ and\ \bibinfo {author}
  {\bibfnamefont {S.}~\bibnamefont {Yokoyama}},\ }\href {\doibase
  10.1088/1475-7516/2005/10/004} {\bibfield  {journal} {\bibinfo  {journal}
  {JCAP}\ }\textbf {\bibinfo {volume} {0510}},\ \bibinfo {pages} {004}
  (\bibinfo {year} {2005})},\ \Eprint {http://arxiv.org/abs/astro-ph/0506262}
  {arXiv:astro-ph/0506262 [astro-ph]} \BibitemShut {NoStop}%
%%CITATION = ASTRO-PH/0506262;%%
\bibitem [{\citenamefont {Matsuda}(2009)}]{Matsuda:2009kp}%
  \BibitemOpen
  \bibfield  {author} {\bibinfo {author} {\bibfnamefont {T.}~\bibnamefont
  {Matsuda}},\ }\href {\doibase 10.1016/j.physletb.2009.11.001} {\bibfield
  {journal} {\bibinfo  {journal} {Phys. Lett.}\ }\textbf {\bibinfo {volume}
  {B682}},\ \bibinfo {pages} {163} (\bibinfo {year} {2009})},\ \Eprint
  {http://arxiv.org/abs/0906.2525} {arXiv:0906.2525 [hep-th]} \BibitemShut
  {NoStop}%
%%CITATION = ARXIV:0906.2525;%%
\bibitem [{\citenamefont {Watanabe}(2012)}]{PhysRevD.85.103505}%
  \BibitemOpen
  \bibfield  {author} {\bibinfo {author} {\bibfnamefont {Y.}~\bibnamefont
  {Watanabe}},\ }\href {\doibase 10.1103/PhysRevD.85.103505} {\bibfield
  {journal} {\bibinfo  {journal} {Phys. Rev. D}\ }\textbf {\bibinfo {volume}
  {85}},\ \bibinfo {pages} {103505} (\bibinfo {year} {2012})}\BibitemShut
  {NoStop}%
\bibitem [{\citenamefont {Abolhasani}\ \emph {et~al.}(2013)\citenamefont
  {Abolhasani}, \citenamefont {Emami}, \citenamefont {Firouzjaee},\ and\
  \citenamefont {Firouzjahi}}]{Abolhasani:2013zya}%
  \BibitemOpen
  \bibfield  {author} {\bibinfo {author} {\bibfnamefont {A.~A.}\ \bibnamefont
  {Abolhasani}}, \bibinfo {author} {\bibfnamefont {R.}~\bibnamefont {Emami}},
  \bibinfo {author} {\bibfnamefont {J.~T.}\ \bibnamefont {Firouzjaee}}, \ and\
  \bibinfo {author} {\bibfnamefont {H.}~\bibnamefont {Firouzjahi}},\ }\href
  {\doibase 10.1088/1475-7516/2013/08/016} {\bibfield  {journal} {\bibinfo
  {journal} {JCAP}\ }\textbf {\bibinfo {volume} {1308}},\ \bibinfo {pages}
  {016} (\bibinfo {year} {2013})},\ \Eprint {http://arxiv.org/abs/1302.6986}
  {arXiv:1302.6986 [astro-ph.CO]} \BibitemShut {NoStop}%
%%CITATION = ARXIV:1302.6986;%%
\bibitem [{\citenamefont {Talebian-Ashkezari}\ \emph
  {et~al.}(2018)\citenamefont {Talebian-Ashkezari}, \citenamefont {Ahmadi},\
  and\ \citenamefont {Abolhasani}}]{Talebian-Ashkezari:2016llx}%
  \BibitemOpen
  \bibfield  {author} {\bibinfo {author} {\bibfnamefont {A.}~\bibnamefont
  {Talebian-Ashkezari}}, \bibinfo {author} {\bibfnamefont {N.}~\bibnamefont
  {Ahmadi}}, \ and\ \bibinfo {author} {\bibfnamefont {A.~A.}\ \bibnamefont
  {Abolhasani}},\ }\href {\doibase 10.1088/1475-7516/2018/03/001} {\bibfield
  {journal} {\bibinfo  {journal} {JCAP}\ }\textbf {\bibinfo {volume} {1803}},\
  \bibinfo {pages} {001} (\bibinfo {year} {2018})},\ \Eprint
  {http://arxiv.org/abs/1609.05893} {arXiv:1609.05893 [gr-qc]} \BibitemShut
  {NoStop}%
%%CITATION = ARXIV:1609.05893;%%
\bibitem [{\citenamefont {Romano}\ \emph
  {et~al.}(2016{\natexlab{b}})\citenamefont {Romano}, \citenamefont {Mooij},\
  and\ \citenamefont {Sasaki}}]{Romano:2015vxz}%
  \BibitemOpen
  \bibfield  {author} {\bibinfo {author} {\bibfnamefont {A.~E.}\ \bibnamefont
  {Romano}}, \bibinfo {author} {\bibfnamefont {S.}~\bibnamefont {Mooij}}, \
  and\ \bibinfo {author} {\bibfnamefont {M.}~\bibnamefont {Sasaki}},\ }\href
  {\doibase 10.1016/j.physletb.2016.02.054} {\bibfield  {journal} {\bibinfo
  {journal} {Phys. Lett.}\ }\textbf {\bibinfo {volume} {B755}},\ \bibinfo
  {pages} {464} (\bibinfo {year} {2016}{\natexlab{b}})},\ \Eprint
  {http://arxiv.org/abs/1512.05757} {arXiv:1512.05757 [gr-qc]} \BibitemShut
  {NoStop}%
%%CITATION = ARXIV:1512.05757;%%
\bibitem [{\citenamefont {Maldacena}(2003)}]{Maldacena:2002vr}%
  \BibitemOpen
  \bibfield  {author} {\bibinfo {author} {\bibfnamefont {J.~M.}\ \bibnamefont
  {Maldacena}},\ }\href {\doibase 10.1088/1126-6708/2003/05/013} {\bibfield
  {journal} {\bibinfo  {journal} {JHEP}\ }\textbf {\bibinfo {volume} {05}},\
  \bibinfo {pages} {013} (\bibinfo {year} {2003})},\ \Eprint
  {http://arxiv.org/abs/astro-ph/0210603} {arXiv:astro-ph/0210603 [astro-ph]}
  \BibitemShut {NoStop}%
%%CITATION = ASTRO-PH/0210603;%%
\bibitem [{\citenamefont {Creminelli}\ and\ \citenamefont
  {Zaldarriaga}(2004)}]{Creminelli:2004yq}%
  \BibitemOpen
  \bibfield  {author} {\bibinfo {author} {\bibfnamefont {P.}~\bibnamefont
  {Creminelli}}\ and\ \bibinfo {author} {\bibfnamefont {M.}~\bibnamefont
  {Zaldarriaga}},\ }\href {\doibase 10.1088/1475-7516/2004/10/006} {\bibfield
  {journal} {\bibinfo  {journal} {JCAP}\ }\textbf {\bibinfo {volume} {0410}},\
  \bibinfo {pages} {006} (\bibinfo {year} {2004})},\ \Eprint
  {http://arxiv.org/abs/astro-ph/0407059} {arXiv:astro-ph/0407059 [astro-ph]}
  \BibitemShut {NoStop}%
%%CITATION = ASTRO-PH/0407059;%%
\end{thebibliography}%

 \end{document}